\documentclass[pra,twocolumn,10pt,aps,showpacs,floatfix,amsmath,amsfonts,letterpaper,longbibliography]{revtex4-1}

\usepackage{stix} % Use Times fonts
\usepackage{graphicx}
\usepackage{amsmath}
\usepackage{url}
\usepackage{array}
\usepackage{multirow}
\usepackage{bm}

\newcolumntype{C}{>{\centering\arraybackslash}p{2.5cm}}

\makeatletter
\def\beq{\@ifstar{\@ifnextchar[{\@beqslabel}{\@beqsnolabel}}
{\@ifnextchar[{\@beqlabel}{\@beqnolabel}}}
\def\@beqlabel[#1]{\begin{equation}\label{#1}}
\def\@beqnolabel{\begin{equation}}
\def\@beqslabel[#1]{\begin{equation*}\label{#1}}
\def\@beqsnolabel{\begin{equation*}}
\def\eeq{\@ifstar{\end{equation*}}{\end{equation}}}
\makeatother
\newcommand{\refcite}[1]{Ref.~\cite{#1}}

\newcommand{\refeq}[1]{Eq.~(\ref{#1})}
\newcommand{\refeqs}[2]{Eqs.~(\ref{#1})\nobreakdash--(\ref{#2})}
\newcommand{\refeqand}[2]{Eqs.~(\ref{#1}) and (\ref{#2})}

\newcommand{\reffig}[1]{Fig.~\ref{#1}}

\newcommand{\refsec}[1]{Sec.~\ref{#1}}

\newcommand{\reftab}[1]{Tab.~\ref{#1}}

\newcommand{\ee}{\mathrm{e}}
\newcommand{\ii}{\mathrm{i}}

\newcommand{\del}{\bm{\nabla}}
\newcommand{\sigmam}{\bm{\sigma}}
\newcommand{\punc}[1]{\,{\text{#1}}}
\newcommand{\sub}[1]{_{\text{#1}}}

\newcommand{\spr}[1]{^{(#1)}}

\newcommand{\bigblp}{\bm{\big(}}
\newcommand{\bigbrp}{\bm{\big)}}
\newcommand{\rv}{\bm{r}}
\newcommand{\Rv}{\bm{R}}
\newcommand{\deltav}{\bm{\delta}}
\newcommand{\goC}{\mathfrak{C}}
\newcommand{\Mv}{\bm{M}}
\newcommand{\zerov}{\bm{0}}
\newcommand{\Phiv}{\bm{\Phi}}
\newcommand{\Bv}{\bm{B}}
\newcommand{\Av}{\bm{A}}
\newcommand{\scL}{\mathcal{L}}
\newcommand{\zv}{\bm{z}}

\DeclareMathOperator{\var}{Var}

\usepackage[usenames,dvipsnames]{color}

\begin{document}

\title{Synchronization transition in the double dimer model on the cubic lattice}

\author{Neil Wilkins}
\affiliation{School of Physics and Astronomy, The University of Nottingham, Nottingham, NG7 2RD, United Kingdom}

\author{Stephen Powell}
\affiliation{School of Physics and Astronomy, The University of Nottingham, Nottingham, NG7 2RD, United Kingdom}

\begin{abstract}
We study the classical cubic-lattice double dimer model, consisting of two coupled replicas of the close-packed dimer model, using a combination of theoretical arguments and Monte Carlo simulations. Our results establish the presence of a `synchronization transition' at a critical value of the coupling, where both replicas remain disordered but their fluctuations become strongly correlated. We show that this unconventional transition, which has neither external nor spontaneous symmetry breaking, is continuous and belongs to the 3D inverted-XY universality class. By adding aligning interactions for dimers within each replica, we map out the full phase diagram including the interplay between columnar ordering and synchronization. We also solve the coupled double dimer model exactly on the Bethe lattice and show that it correctly reproduces the qualitative phase structure, but with mean-field critical behavior.
\end{abstract}

\maketitle

\section{Introduction}
\label{introduction}

Confinement phase transitions bring together a number of important concepts in modern condensed matter physics. These include fractionalized excitations \cite{Balents2010,Henley2010}, i.e., emergent objects that cannot be constructed from finite combinations of the elementary constituents; non-Landau phase transitions \cite{Senthil2004,Senthil2004b}, where a description in terms of a symmetry-breaking order parameter is not sufficient to describe the critical behavior; and topological order \cite{Castelnovo2012}, loosely defined as structure that is detectable only through global measurements.

Such transitions are known to exist in classical dimer models \cite{Kasteleyn1961,Temperley1961}, in which the degrees of freedom are \emph{dimers}, objects that occupy the bonds of a lattice subject to a close-packing constraint: Each site is touched by exactly one dimer. Introducing a pair of empty sites -- defects in the constraint referred to as \emph{monomers} -- one can distinguish phases by the effective interactions induced between the monomers by the background dimer configurations. A confinement transition separates a phase where these interactions are confining, i.e., the potential grows without bound with increasing distance, from one where they are not, and so the monomers can be separated to infinity with finite free-energy cost \cite{Henley2010}.

Known examples of confinement transitions in classical dimer models can be divided into two types. The first, where confinement occurs simultaneously with spontaneous symmetry breaking, includes the well-studied case of the columnar-ordering transition in the dimer model on the cubic lattice \cite{Alet2006} and the corresponding square-lattice transition \cite{Alet2005,Alet2006b}. The second type is where symmetry is broken externally, and includes the Kasteleyn transition in the honeycomb-lattice dimer model \cite{Kasteleyn1963} and its generalizations to three dimensions (3D) \cite{Bhattacharjee1983,Jaubert2008}, as well as the `1GS' variant of the cubic dimer model studied by Chen et al. \cite{Chen2009}.

Here, we demonstrate the existence of a `pure' confinement transition, with neither spontaneous nor external symmetry breaking, in the \emph{double dimer model} \cite{Raghavan1997,Damle2012,Kenyon2014}, comprising two coupled replicas of the close-packed dimer model. In this paper we focus on the cubic lattice, and confirm  using Monte Carlo (MC) simulations the presence of a synchronization transition at a finite ratio of temperature \(T\) to the coupling between replicas. Subsequent work will address the case of the square lattice.

\begin{figure}[b]
\includegraphics[width=\columnwidth]{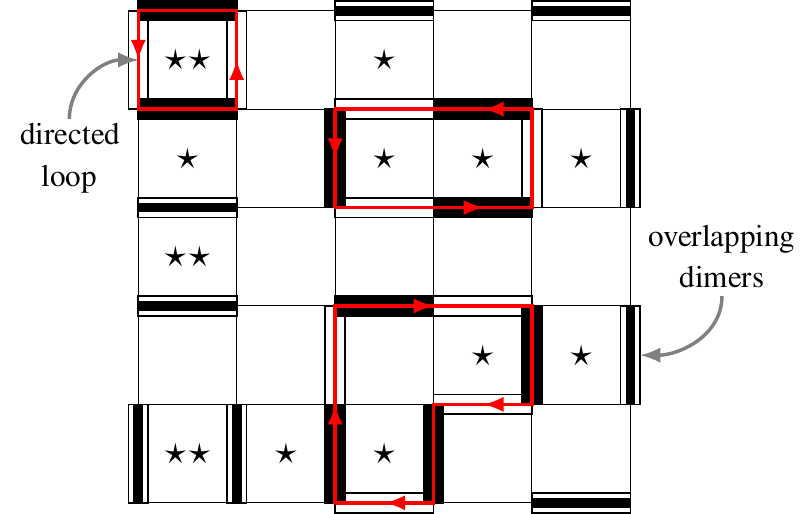}
\caption{An example configuration of the double dimer model, in which two replicas of the close-packed dimer model (shown in black and white) are defined on the same lattice. Although we consider a cubic lattice, illustrations are shown on the square lattice for clarity. If overlapping dimers are deleted, the result is a gas of directed loops (red). According to \refeq{eq:configurationenergy}, parallel pairs of nearest-neighbour dimers within each replica (marked with a star, \(\star\)) contribute $+J$ to the energy, and overlapping dimers contribute $+K$ to the energy. Hence, the energy of this configuration is $E = 15J + 9K$.}
\label{fig:model}
\end{figure}
To understand the transition, consider overlaying any pair of dimer configurations and deleting all dimers that coincide. As illustrated in \reffig{fig:model}, the result is a gas of directed loops, referred to as the `transition graph' and corresponding to the set of dimer rearrangements that take one configuration to the other \cite{Kasteleyn1963,Krauth2003}. A coupling that favors overlapping dimers then amounts to an energy cost per unit loop length. This makes possible a loop proliferation (or `condensation') transition, between a phase at low \(T\) with only sparse short loops and one at high \(T\) with a finite density of boundary-spanning loops, as a result of competition between energy and entropy \cite{Banks1977,Dasgupta1981}. (See \refsec{model} for a precise definition.) In terms of the dimers, the proliferation of loops amounts to a phase transition between replicas being mostly synchronized and mostly independent, which we will refer to as a `synchronization transition'.

An equivalent characterization is provided by the concept of confinement: Consider first removing a single dimer in one replica, leaving an adjacent pair of monomers, and then rearranging dimers to separate the monomers to arbitrary distance \(R\). In terms of the overlap between the two replicas, this necessarily produces an open string joining the pair. The condition that loops proliferate is equivalent to the entropy of a long string overcoming its energy cost (\(\propto R\)) \cite{Savit1980} and hence that such a monomer pair is deconfined \cite{Henley2010}. Note that a double pair of monomers, one in each replica, is instead deconfined in both phases, including in the limit of perfect synchronization between the layers.

As is apparent from the nature of the two phases, the synchronization transition on the cubic lattice is an unusual example of a classical transition in 3D with no local order parameter. We argue theoretically that the transition, if continuous, belongs in the inverted-XY universality class, and demonstrate using MC results that this is indeed the case. It can therefore be seen as an unusual example of a scalar Higgs transition, similar to the 1GS model \cite{Chen2009} and the helical-field transition in spin ice \cite{Powell2012,Powell2013}, but without any external symmetry breaking.

Previous studies of the double dimer model have focused on the square lattice, and particularly on the cases where the coupling between replicas is either infinite \cite{Raghavan1997} or vanishing \cite{Kenyon2014}. Damle et al.\ \cite{Damle2012} have showed that certain matrix elements in a \emph{quantum} dimer model can be related to the partition function of an interacting double dimer model.

A related transition between low- and high-overlap phases has been observed in paired replicas of plaquette spin models \cite{Garrahan2014,Turner2015}. While there are many similarities between these transitions, it should be noted that the plaquette models have no analogue of the confinement criterion, and the boundary is a first-order line terminating at a critical endpoint. Recent work on the bilayer Kitaev model \cite{Tomishige2018} has also observed a transition from a spin liquid to a phase where spins on the two layers pair into singlets.

\subsection*{Outline}

In \refsec{model} we define the double dimer model, including couplings between and within replicas, and present theoretical arguments for its phase structure and critical properties. We then describe, in \refsec{wormalgorithm}, the MC method that we use, which extends the standard worm algorithm. Our numerical results, including a detailed study of the critical properties of the synchronization transition, are presented \refsec{numericalresults}. Finally, in \refsec{bethelattice}, we show that the double dimer model, including coupling between replicas, can be solved exactly on the Bethe lattice. We conclude in \refsec{sec:conclusions} with a brief discussion of analogous synchronization transitions in other systems and of possible experimental realizations.

\section{Model}
\label{model}

We consider the double dimer model, consisting of two replicas $\alpha \in \{1,2\}$ of the dimer model, both defined on an \(L\times L\times L\) cubic lattice with periodic boundaries. Denoting by $d_l\spr{\alpha}$ the dimer occupation number (equal to zero or one) of replica \(\alpha\) on bond $l$, the configuration energy can be written as
\beq[eq:configurationenergy]
E = J\left[N_{\parallel}\spr{1}+N_{\parallel}\spr{2}\right]+K\sum_{l} d_{l}\spr{1} d_{l}\spr{2}\punc,
\eeq
where $N_{\parallel}\spr{\alpha}$ denotes the number of parallel pairs of nearest-neighbour dimers \cite{Alet2005,Alet2006b, Alet2006} in replica \(\alpha\). An example configuration is shown in \reffig{fig:model}.

\begin{figure*}
\includegraphics[width=\textwidth]{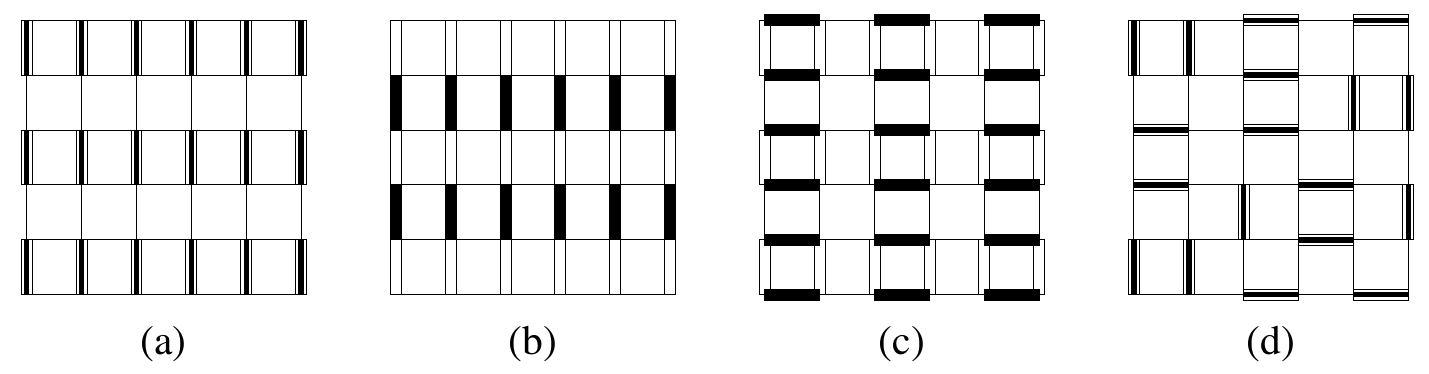}
\caption{Example ground states of the double dimer model of \refeq{eq:configurationenergy}, illustrated for the square lattice but applying also to the cubic lattice. (a)--(c) Columnar configurations, which minimize the energy for $J<0$, $K=0$. In each replica \(\alpha \in \{1,2\}\) (white and black dimers, respectively), the dimers are arranged in columns, maximizing the number of parallel plaquettes \(N_\parallel\spr\alpha\) and hence minimizing the energy. For \(K=0\), the two replicas are uncoupled and so all three arrangements have equal energy. For $J<0$, $K<0$ configuration (a), the columnar \& synchronized ground state, maximizes the overlap and hence minimizes the energy. Configurations (b) and (c) are columnar \& antisynchronized, with replica magnetizations antiparallel and perpendicular respectively; both have zero overlap between replicas and so are degenerate ground states for $J<0$, $K>0$.
(d) Example of a fully synchronized configuration, one of an extensive number of ground states for $J=0$, $K<0$. Each replica is disordered, but the overlap between their configurations is maximal, minimizing the energy.}
\label{fig:groundstates}
\end{figure*}

When $J<0$, the first term favors parallel alignment of dimers within each replica. It is minimized by configurations with columnar order, as illustrated in Figs.~\ref{fig:groundstates}(a)--(c), which break translation and rotation symmetries. A suitable order parameter is the `magnetization' \(\Mv\spr{\alpha}\), a vector for each replica \(\alpha\) with components
\beq[eq:magnetization]
M_{\mu}\spr{\alpha} = \frac{2}{L^{3}}\sum_{\rv}(-1)^{r_{\mu}}d_{\rv,\mu}\spr{\alpha}
\punc,
\eeq
where \(\rv,\mu\) denotes the bond between sites \(\rv\) and \(\rv + \deltav_\mu\) (with \(\deltav_\mu\) a unit vector in direction \(\mu \in \{x,y,z\}\)). On the cubic lattice, there are six such configurations for each replica with \(\Mv\spr{\alpha} = \pm \deltav_\mu\).

The second term in \refeq{eq:configurationenergy} couples the two replicas, by counting the number of bonds that are occupied in both. We mainly focus on the case \(K < 0\), where the effect is to favor overlapping dimers; a fully synchronized example, minimizing this term, is shown in \reffig{fig:groundstates}(d).

The dimer configurations are subject to the close-packing constraint \(n_{\rv}\spr{\alpha} = 1\), where
\beq
n_{\rv}\spr{\alpha} = \sum_\mu \left[ d_{\rv,\mu}\spr{\alpha} + d_{\rv-\deltav_\mu,\mu}\spr{\alpha}\right]
\eeq
is the number of dimers at site \(\rv\). This constraint is applied separately to each replica \(\alpha\), and so the partition function
\beq[eq:partitionfunction]
Z = \sum_{\substack{c\spr{1}\in\goC_0\\c\spr{2}\in\goC_0}} \ee^{-E/T}
\eeq
sums, for each \(\alpha\), over the set \(\goC_0\) of all configurations that obey \(n_{\rv}\spr{\alpha} = 1\) for all \(\rv\). (We set \(k\sub{B} = 1\).)

\subsection{Loop picture}
\label{looppicture}

On a bipartite lattice, such as the cubic lattice, configurations may be represented by a `magnetic field', defined on the bonds of the lattice \cite{Huse2003,Henley2010}. Specifically, bond \(\rv,\mu\) is assigned the field
\beq[eq:magneticfield]
B_{\bm{r},\mu}\spr{\alpha} = \epsilon_{\rv}\left[d_{\rv,\mu}\spr{\alpha}-\frac{1}{q}\right]\punc,
\eeq
where $\epsilon_{\rv} = (-1)^{r_x + r_y + r_z} = \pm 1$ depending on the sublattice and $q$ is the coordination number.

The close-packing constraint for the dimers is equivalent to the condition that the (magnetic) `charge', given by the lattice divergence of \(B_{\rv,\mu}\spr{\alpha}\),
\beq[eq:defineQ]
Q\spr{\alpha}_{\rv} = \sum_\mu \left[ B_{\rv,\mu}\spr{\alpha} - B_{\rv-\deltav_\mu,\mu}\spr{\alpha} \right]\punc,
\eeq
is zero on every site. The normalization of \(B_{\bm{r},\mu}\spr{\alpha}\) is chosen so that removing a dimer (and so breaking the close-packing constraint) leaves a pair of monomers on opposite sublattices with \(Q_{\rv} = \pm 1\).

When the two replicas are overlaid, the result can be interpreted as a set of directed loops. To see this, consider the relative magnetic field \(B_{\rv,\mu}\spr{-} = B_{\rv,\mu}\spr{1} - B_{\rv,\mu}\spr{2}\), which takes values on each bond of \(\pm 1\) or \(0\). The former is interpreted as a loop element directed along \(\pm \deltav_\mu\), while the latter means that the dimers overlap and is interpreted as the absence of a loop element. Since the relative field is clearly also divergenceless, these elements indeed form a set of closed loops. Note that swapping the two replicas switches the direction of each loop.

Given the magnetic field \(B_{\rv,\mu}\spr{\alpha}\), it is possible to define a vector flux \(\Phiv\spr{\alpha}\) by
\begin{align}
\Phi_{\mu}\spr{\alpha} &= \frac{1}{L}\sum_{\rv}B_{\rv,\mu}\spr{\alpha} \\ &= \frac{1}{L}\sum_{\rv}\epsilon_{\bm{r}}d_{\rv,\mu}\spr{\alpha}
\punc,
\end{align}
which is equivalent to the sum of the magnetic fields on links crossing a surface normal to \(\deltav_\mu\). It is therefore invariant under local dimer rearrangements; in fact, with periodic boundary conditions (PBCs), changes in flux are only possible by shifting dimers around a loop encircling the whole system. For example, adding a loop in \(B_{\rv,\mu}\spr{-}\) that spans the system once in direction \(\mu\) increases the relative flux \(\Phi_\mu\spr{-} = \Phi_\mu\spr{1} - \Phi_\mu\spr{2}\) by one.

\subsection{Phase diagram}
\label{phasediagram}

\begin{figure}
\includegraphics[width=\columnwidth]{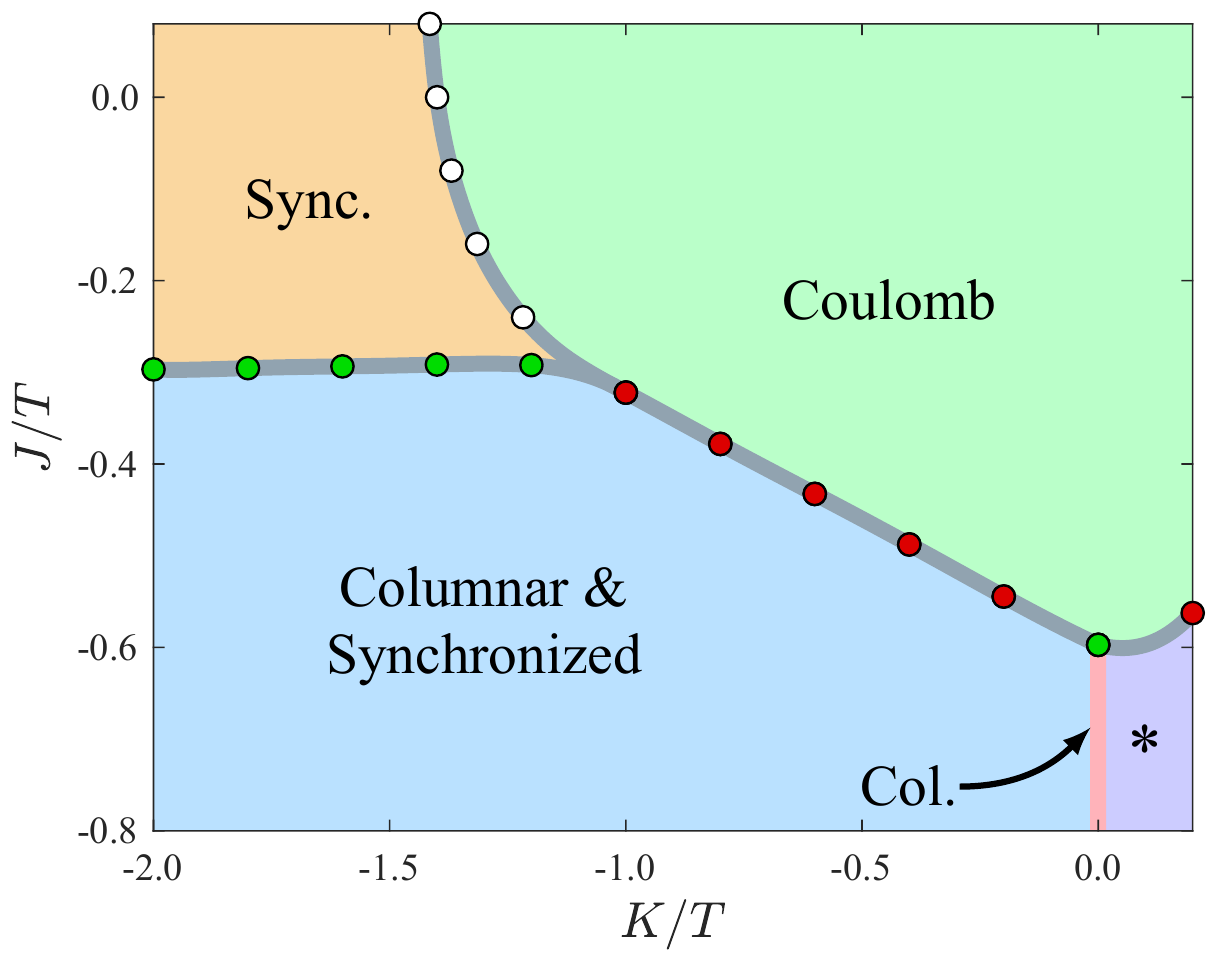}
\caption{Phase diagram for the double dimer model of \refeq{eq:configurationenergy} on the cubic lattice, in the $(J/T,K/T)$ plane. Dots show points where the phase boundary has been determined using MC simulations, and thick grey lines are guides to the eye. The orange region, labeled `Sync.' is the synchronized phase. The pink line labeled `Col.' is the (unsynchronized) columnar phase, known to occur at \(K=0\), whilst the purple region, labeled `$\ast$', is the columnar \& antisynchronized phase. Red dots represent first-order transitions, whilst green and white dots represent continuous transitions, in the tricritical and inverted-XY universality classes, respectively.}
\label{fig:phasediagram}
\end{figure}

The phase diagram of the double dimer model, \refeq{eq:configurationenergy}, obtained using the MC method detailed in \refsec{wormalgorithm}, is shown in \reffig{fig:phasediagram}. In this section we define the phases shown and explain how the phase structure can be understood theoretically. In \refsec{numericalresults} we describe how the phase boundaries, as well as the critical properties at each, are determined in the simulations.

\subsubsection{Independent replicas}
\label{independentreplicas}

We first review the phase structure for \(K=0\), where the two replicas act as independent (single) dimer models. For \(K = J = 0\), the cubic dimer model exhibits a Coulomb phase \cite{Huse2003,Henley2010}, in which the dimers are disordered and their correlations take a dipolar form. A single phase transition at \((J/T)\sub{c}=-0.597\) separates this from a low-temperature phase with nonzero columnar order parameter \(\langle \Mv \rangle \neq \zerov\) \cite{Alet2006}. The transition is apparently continuous with critical exponents compatible with the tricritical universality class. Theoretical arguments \cite{Powell2008,Charrier2008,Chen2009}, however, suggest that the critical properties should be described by the so-called noncompact \(\mathbb{CP}^1\) theory (see \refsec{SecFieldTheories}), and additional interactions \cite{Charrier2010,Papanikolaou2010} indeed modify the exponents to values compatible with this universality class \cite{Sreejith2014}.

Besides the order parameter \(\langle \Mv \rangle\), the Coulomb and columnar phases can be distinguished either through the flux or through the effective interactions between monomers. Consider first the latter, which involves introducing a test pair of monomers with opposite charge into the background of close-packed dimers. In the case of the double dimer model, the pair can be inserted into either replica; we choose replica \(1\), and define the monomer distribution function as
\beq[EqGm]
G\sub{m}(\rv_+ - \rv_-) = \frac{1}{Z}\sum_{\substack{c\spr{1}\in\goC(\rv_+,\rv_-)\\c\spr{2}\in\goC_0}} \ee^{-E/T}\punc.
\eeq
The set \(\goC(\rv_+,\rv_-)\) contains all configurations with charge distribution \(Q_{\rv} = \delta_{\rv,\rv_+} - \delta_{\rv,\rv_-}\), i.e., with monomers of opposite charge at sites \(\rv_\pm\) (e.g., holes on opposite sublattices). The effective interaction \(U\sub{m}\) is defined by \(G\sub{m}(\Rv) = \ee^{-U\sub{m}(\Rv)/T}\), and can be interpreted as the free-energy cost associated with the imposed charge distribution. (Translation symmetry, assuming PBCs, implies \(G\sub{m}\) and \(U\sub{m}\) depend only on the displacement \(\Rv = \rv_+ - \rv_-\).)

For \(K=0\), the replicas are independent, and so \(G\sub{m}\) reduces to the monomer distribution function for the single dimer model. In the Coulomb phase for small \(\lvert J\rvert/T\), \(U\sub{m}\) is a Coulomb potential at large separation, \(U\sub{m}(\rv_+ - \rv_-) \sim U\sub{m}(\infty) - \kappa/(4\pi\lvert\rv_+ - \rv_-\rvert)\), where \(\kappa\) (the `flux stiffness') and \(U\sub{m}(\infty)\) are finite (positive) constants \cite{Henley2010}. In the low-temperature phase, separating the monomers necessarily disrupts the columnar order along a string joining them [consider Figs.~\ref{fig:groundstates}(a)--(c)], and so has a free-energy cost proportional to distance \cite{Henley2010}. This qualitative distinction, between a confining interaction (preventing separation to infinite distance) at low \(T\) and deconfinement at high \(T\), provides an alternative characterization of the phase transition.

In practice, it is convenient to use the confinement length \(\xi\), defined by
\beq[eq:cl]
\xi^{2} = \frac{\sum_{\Rv} \lvert\Rv\rvert^{2}G\sub{m}(\Rv)}{\sum_{\Rv} G\sub{m}(\Rv)}\punc,
\eeq
which represents the root-mean-square separation of the test monomers. In the Coulomb phase, \(G\sub{m}(\Rv) \rightarrow \ee^{-U\sub{m}(\infty)/T} > 0\) for large separation, and so \(\xi \sim L\). In the columnar phase, by contrast, \(U\sub{m}(\Rv)\) grows without limit as \(\lvert\Rv\rvert\rightarrow\infty\), \(G\sub{m}(\Rv)\rightarrow 0\), and so \(\xi\) is an \(L\)-independent constant.

A related criterion for the phases can be expressed in terms of the flux \(\Phiv\spr{\alpha}\). The mean flux vanishes by symmetry in both phases, while the variance, \(\var\Phiv\spr\alpha = \langle\lvert\Phiv\spr\alpha\rvert^2\rangle\), scales differently with system size in the two: In the Coulomb phase, flux fluctuations are large, \(\var\Phiv\spr\alpha \approx L/\kappa\) \cite{Huse2003}. In the columnar phase, the variance is exponentially small in \(L\), because shifting dimers along a loop spanning the system disturbs the columnar order and hence costs energy \(E \sim J L\) \footnote{To see the connection to the monomer-confinement criterion, imagine changing the flux by removing a dimer to create a monomer pair, winding one monomer around the periodic boundaries, and then recombining the pair \cite{Hermele2004}.}. Because the two replicas are independent, the variances of the total and relative flux \(\Phiv\spr\pm = \Phiv\spr1 \pm \Phiv\spr2\) are identical, and equally serve to distinguish the two phases.

\subsubsection{Coupled replicas}
\label{coupledreplicas}

Consider now \(J = 0\) and nonzero coupling \(K<0\) between replicas. In the ensemble defined by \refeq{EqGm}, \(B_{\rv,\mu}\spr{2}\) is divergenceless while \(B_{\rv,\mu}\spr{1}\) has nonzero divergence at \(\rv_+\) and \(\rv_-\). This implies the presence of an open string in \(B_{\rv,\mu}\spr{-}\) that runs between these two sites, and along which the two replicas differ. In the limit \(K/T \rightarrow -\infty\), the string will take the shortest possible path, resulting in an energy proportional to separation and hence a confining effective interaction \(U\sub{m}\). Comparing this limit with the case where \(K=0\) (and \(J=0\)), it follows that there must be a confinement transition, a qualitative change in the large-separation form of \(U\sub{m}\), between the two. In our results, shown in \reffig{fig:phasediagram}, we indeed find such a transition at a critical coupling \((K/T)\sub{c}=-1.400\).

At temperatures above this point, where the entropy of the open string overcomes its energy cost, closed loops of \(B_{\rv,\mu}\spr{-}\) are also free-energetically favorable. As a result, loops spanning the system boundaries, which cost an energy \(E\sim KL\) and are hence suppressed exponentially in \(L\) at low temperatures, `proliferate' at the transition. Because these loops change the relative flux \(\Phiv\spr-\), the high-temperature phase has \(\var\Phiv\spr- \approx L/\kappa_-\), as at \(K=0\) but with a modified flux stiffness \(\kappa_-\). By contrast, the variance of the total flux, \(\var\Phiv\spr{+}\), is large (\(\approx L/\kappa_+\)) in both phases, because identical loops in both replicas costs zero energy [consider \reffig{fig:groundstates}(d)]. The behavior of the flux variances in the different phases is summarized in \reftab{tab:phases}.
\renewcommand{\arraystretch}{1.2}
\begin{table*}
\begin{center}
\begin{tabular}{ | C | C | C | C | C | }
\hline
 & Coulomb & Columnar & Synchronized\\ \hline
\(\var\Phiv\spr-\) & Large &  Small & Small\\ \hline
\(\var\Phiv\spr+\) & Large &  Small & Large\\ \hline
\(\xi\) & Large & Small & Small\\
\hline
\end{tabular}
\caption{Behavior in different phases of the double dimer model of: the variance of the flux difference, \(\Phiv\spr- = \Phiv\spr1-\Phiv\spr2\); the variance of the total flux, \(\Phiv\spr+ = \Phiv\spr1+\Phiv\spr2\); and the confinement length, \(\xi\). `Small' means \(\var\Phiv\spr\pm\) decreases exponentially with linear system size \(L\) and \(\xi\) is independent of \(L\), while `Large' means \(\var\Phiv\spr\pm\sim L\) and \(\xi \sim L\). In the columnar \& (anti)synchronized phases, these observables behave as in the columnar phase.}
\label{tab:phases}
\end{center}
\end{table*}
\renewcommand{\arraystretch}{1}

While confinement and the flux variance thereby provide precise definitions of the phases, we also expect loop proliferation to reduce the overlap \(\sum_l d_l\spr{1}d_l\spr{2}\) between the replicas. We therefore refer to the low-\(T\), high-overlap phase as `synchronized' and the high-\(T\), low-overlap phase as `unsynchronized', although the overlap is nonzero in both phases and so does not provide an order parameter in the strict sense.

It should be noted that the energy \(\propto K\) associated with each element of a directed loop (or open string) is not the only contribution to its free-energy cost. In regions devoid of loops, overlapping dimers can be rearranged without changing the loop configuration. (For example, in the configuration of \reffig{fig:model}, flipping the parallel pair of overlapping dimers around the bottom-left plaquette in both replicas does not create a new directed loop). To leading order, this results in an entropy that scales with the number of overlapping dimers. Since the introduction of a directed loop reduces this number, and so the entropy, at finite \(K/T\) we expect an additional free-energy cost per unit length of loops, which can be thought of as renormalizing \(K/T\) towards more negative values \footnote{The significance of this effect can be estimated by comparing with a simple approximation that neglects it and treats the loops as simply costing energy \({\frac{1}{2}\lvert K\rvert}\) per unit length. (A loop of length \(\ell\) reduces the number of overlapping dimers by \({\frac{1}{2}\ell}\).) This model has a proliferation transition at \({T\sub{c} \simeq 0.33\lvert K\rvert}\) \cite{Dasgupta1981}. In fact, our MC simulations give \({T\sub{c} = 0.714\lvert K\rvert}\) (see \reffig{fig:tx}) -- the additional free-energy cost of loops due to the entropy of overlapping dimers, neglected in our approximation, means a higher temperature is needed for them to proliferate.}.

The arguments for the phase structure can be straightforwardly extended to include both \(J/T\) and \(K/T\). At large negative \(J/T\) and \(K = 0\), both replicas are columnar ordered, but the relative orientations of the two magnetizations \(\Mv\spr{1}\) and \(\Mv\spr{2}\) are arbitrary. Infinitesimal negative \(K/T\) is sufficient to split this degeneracy extensively and therefore to synchronize the two replicas, giving the columnar \& synchronized phase, illustrated in \reffig{fig:groundstates}(a), with \(\langle\Mv\spr{1}\rangle=\langle\Mv\spr{2}\rangle\).

For positive \(K/T\), any pair of columnar configurations with distinct magnetizations has zero overlap and hence minimal energy. There is an accidental degeneracy, between antiparallel (\(\Mv\spr{1} = -\Mv\spr{2}\)) and perpendicular (\(\Mv\spr{1}\cdot\Mv\spr{2} = 0\)) magnetizations in the two replicas, illustrated in Figs.~\ref{fig:groundstates}(b) and (c) respectively, which can be resolved by order by disorder \cite{Henley1989,Chalker2017}. The elementary fluctuations, which involve flipping a single pair of parallel dimers around a plaquette, cost energy \(+6\lvert J\rvert\) in both cases, but additionally may cost energy \(+2K\) in the case of perpendicular magnetization. The free energy is therefore lower in the antiparallel arrangement, suggesting that this is selected by order by disorder. Our MC results (see \refsec{columnar&synchronized-columnar&antisynchronized}) are indeed consistent with a phase where \(\langle\Mv\spr{1}\rangle = -\langle\Mv\spr{2}\rangle\), which we refer to as columnar \& antisynchronized.

Comparison with the single dimer model further allows some quantitative details of the phase boundaries to be inferred: The critical point separating the Coulomb and columnar phases for \(K=0\) is clearly \((J/T)\sub{c} = -0.597\) as for the single dimer model. Similarly, when \(K/T\rightarrow -\infty\), the two replicas are perfectly synchronized, and behave as a single dimer model with effective interaction \(J\sub{eff}=2J\). The critical temperature for columnar ordering is therefore given by \(\frac{1}{2}(J/T)\sub{c} = -0.299\) in this limit. As shown in \reffig{fig:phasediagram}, the critical value of \(J/T\) closely approximates this limiting value already for \(K/T = -2\).

\subsection{Field theories and critical properties}
\label{SecFieldTheories}

A continuum description for the Coulomb phase in the single dimer model is given by replacing the effective magnetic field \(B_{\rv,\mu}\) by a continuum vector field \(\Bv\) \cite{Huse2003,Henley2010}. The latter is subject to the constraint \(\del\cdot\Bv = 0\), inherited from the close-packing constraint on the dimers, and hence can be expressed as \(\Bv = \del\times\Av\) in terms of the vector potential \(\Av\). The continuum (Euclidean) action density is then given by
\beq
\scL\sub{SDM} = \frac{1}{2}\kappa \lvert\Bv\rvert^2 = \frac{1}{2}\kappa \lvert \del \times \Av \rvert^2\punc,
\eeq
where \(\kappa\) is the flux stiffness introduced in \refsec{independentreplicas}, plus irrelevant higher-order terms.

In the double dimer model, one can similarly introduce a continuum magnetic field \(\Bv\spr\alpha\) for each replica, with the same stiffness \(\kappa\) for each. The coupling \(K\) leads to a term \(\lambda \Bv\spr{1}\cdot \Bv\spr{2}\), with \(\lambda \sim K\), and so an effective action for the unsynchronized Coulomb phase can be written as
\beq
\scL\sub{DDM} = \frac{1}{2}\kappa_+ \lvert \Bv\spr+ \rvert^2 + \frac{1}{2}\kappa_- \lvert \Bv\spr- \rvert^2\punc,
\eeq
where \(\Bv\spr\pm = \Bv\spr1\pm\Bv\spr2\) and \(\kappa_\pm = \frac{1}{2}(\kappa \pm \lambda)\). The synchronization transition, at which fluctuations of \(\Bv\spr-\) are suppressed, occurs when \(K < 0\) and hence \(\kappa_- > \kappa_+\).

Confinement transitions from the Coulomb phase, such as the synchronization and columnar-ordering transitions, can be described by introducing `matter' fields to enforce the restriction to discrete values \cite{Chen2009,Powell2011}. Condensation of these fields then leads, by the Higgs mechanism, to suppression of magnetic-field fluctuations. The structure of the critical theory is determined by considering representations of the projective symmetry group (PSG) \cite{Wen2002} under which the matter fields transform.

In the case of the columnar-ordering transition in the single dimer model, the critical theory is \cite{Powell2008,Charrier2008,Chen2009}
\beq
\scL\sub{SDM,crit.} = \scL\sub{SDM} + \lvert (\del - \ii \Av) \zv \rvert^2 + s\lvert \zv \rvert^2 + u \lvert \zv \rvert^4 \punc,
\eeq
where \(s\) and \(u\) are real parameters and \(\zv\) is a two-component complex vector (which is said to be `minimally coupled' to \(\Av\)). The PSG analysis shows that the field \(\zv\) transforms as a spinor under real-space rotations and allows one to express the magnetization as \(\Mv \sim \zv^\dagger \sigmam \zv\). In this description, the ordering transition occurs when \(s\) is reduced below its critical value and \(\zv\) condenses, giving a nonzero magnetization and also suppressing fluctuations of the magnetic field via the Higgs mechanism.

In the double dimer model, the matter field should couple identically to both replicas. We therefore expect the critical properties at the columnar-ordering transition in the double dimer model to be the same as in the single-replica case. While some theoretical aspects of this transition remain unresolved \cite{Sreejith2019}, its properties have been well characterized numerically \cite{Alet2006, Charrier2010, Papanikolaou2010}.

To describe the synchronization transition, one must similarly include a matter field \(\varphi\) whose role is to restrict \(B_{\rv,\mu}\spr-\) to \(\pm 1\) or \(0\). Because these values are integers, the PSG is trivial in this case \footnote{In the notation of \refcite{Chen2009}, the background flux is zero, and so the static gauge configuration \(\bar\alpha\) vanishes.}, and so the result is a scalar Higgs theory,
\beq
\scL\sub{DDM,crit.} = \scL\sub{DDM} + \lvert (\del - \ii \Av\spr-) \varphi \rvert^2 + s_- \lvert \varphi\rvert^2 + u_- \lvert \varphi\vert^4\punc,
\eeq
where \(\Bv\spr- = \del\times\Av\spr-\). We have not included a field coupling to \(\Bv\spr+\), which would remain noncritical across the synchronization transition.

In 3D, the scalar Higgs theory has a continuous transition in the XY universality class but with an inverted temperature axis \cite{Dasgupta1981}. (A more direct route to the same critical theory starts from the loop picture and uses the standard mapping from integer loops to the XY model \cite{Banks1977}.) We therefore expect the synchronization transition to belong to the inverted-XY universality class.

\section{Worm algorithm}
\label{wormalgorithm}

% Introduction
We obtain numerical results using the MC worm algorithm \cite{Sandvik2006}, in which non-local loop updates are performed. With PBCs, loops can span the boundaries, allowing changes in flux.

\subsection{Single loops}
\label{singleloops}

\begin{figure*}
\begin{center}
\includegraphics[width=0.75\textwidth]{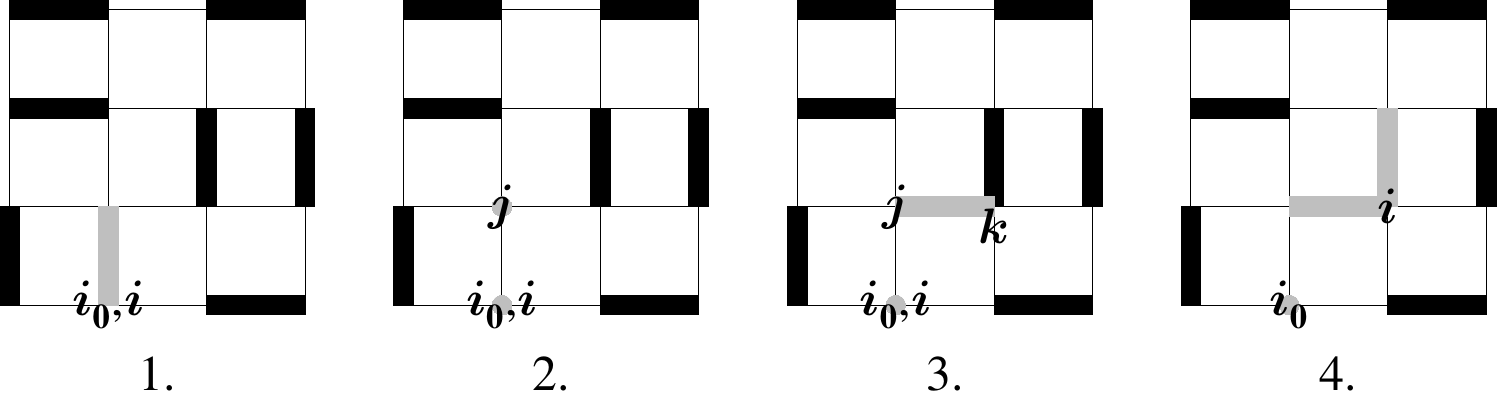}
\caption{Local steps involved in a single-loop update of the worm algorithm (see main text for details).}
\label{fig:steps}
\end{center}
\end{figure*}

% Steps
We begin by reviewing the standard implementation of the worm algorithm, using single loops. As illustrated in \reffig{fig:steps}, the process is broken down into a series of local steps:
\begin{enumerate}
  \item Choose a lattice site $i=i_{0}$ and a replica $\alpha$ at random.
  \item In the current configuration of replica \(\alpha\), site $i$ is connected by a dimer to a neighboring site $j$. Delete this dimer, denoted by $(i,j)^{(\alpha)}$.
  \item Select a neighbour of $j$, called $k$, using a local detailed balance rule (described below), and insert a new dimer $(j,k)^{(\alpha)}$.
  \item If $k=i_{0}$, close the loop. Otherwise return to step 2, using $i=k$.
\end{enumerate}
Since all loops are performed without rejection, the worm algorithm is highly efficient.

% Local detailed balance
The transition probability $P\bigblp(i,j)^{(\alpha)}\rightarrow (j,k)^{(\alpha)}\bigbrp$, with which the site $k$ is selected in step 3, is determined as follows. The requirement for global detailed balance translates into the local detailed balance condition \cite{Sandvik2006}
\begin{multline}
w\bigblp(i,j)^{(\alpha)}\bigbrp P\bigblp(i,j)^{(\alpha)}\rightarrow (j,k)^{(\alpha)}\bigbrp =  \\ w\bigblp(j,k)^{(\alpha)}\bigbrp P\bigblp(j,k)^{(\alpha)}\rightarrow (i,j)^{(\alpha)}\bigbrp\punc.
\end{multline}
Here, $w\bigblp(j,k)^{(\alpha)}\bigbrp$ is the equilibrium probability of the configuration obtained on insertion of the dimer $(j,k)^{(\alpha)}$ in step 3. For the double dimer model, \refeq{eq:configurationenergy} implies
\begin{equation}
w\bigblp(j,k)^{(\alpha)}\bigbrp \propto \exp\left[-\left(JN_{(j,k)}^{(\alpha)}+Kd_{(j,k)}^{(\bar\alpha)}\right)/T\right],
\end{equation}
where $N_{(j,k)}^{(\alpha)} \in \{0,1,2,3,4\}$ is the number of nearest-neighbour dimers parallel to $(j,k)^{(\alpha)}$ in the same replica, whilst $d_{(j,k)}^{(\bar\alpha)}$ is the dimer occupation number of the other replica \(\bar\alpha\) on bond $(j,k)$. A solution for the transition probabilities, chosen to reduce backtracks (where \(k=i\)), is then
\begin{equation}
\label{eq:tp}
P\bigblp(i,j)^{(\alpha)}\rightarrow(j,k)^{(\alpha)}\bigbrp = \frac{w\bigblp(j,k)^{(\alpha)}\bigbrp-\textrm{min}(\bm{w})\delta_{i,k}}{\sum_{k}w\bigblp(j,k)^{(\alpha)}\bigbrp-\textrm{min}(\bm{w})},
\end{equation}
where $\bm{w}$ is a $6$-component vector containing elements $w\bigblp(j,k)^{(\alpha)}\bigbrp$ for all \(k\) \cite{Sreejith2014}.

% Monomer distribution function
Step 2 produces configurations containing two test monomers: a stationary monomer at site $i_{0}$, and a moving monomer at site $j$ (see \reffig{fig:steps}). Hence, single-loop updates may be used to construct the monomer distribution function $G\sub{m}(\Rv)$, by tallying the monomer separation $\Rv$ after each step 2. Since the local detailed balance rule correctly samples only configurations produced by step 3, it is necessary to tally an amount $1/\sum_{k}w\bigblp(j,k)^{(\alpha)}\bigbrp$, rather than unity \cite{Alet2006b}.

\subsection{Double loops}
\label{doubleloops}

% Motivation
As discussed in \refsec{coupledreplicas}, single loops are suppressed in the synchronized phase, whereas simultaneous loops in both replicas are not. Therefore, to avoid problems with ergodicity, it is necessary to perform double-loop updates.

% Implementation
Double loops are constructed as follows. In step 1, loops begin in both replicas from the same randomly chosen site $i=l=i_{0}$. Both loops perform step 2 as for a single-loop update, deleting dimers $(i,j)^{(1)}$ and $(l,m)^{(2)}$. Step 3 now corresponds to $36$ choices, with $6$ in each replica. The insertion of new dimers $(j,k)^{(1)}$ and $(m,n)^{(2)}$ is associated with a configuration probability
\begin{multline}
\label{eq:weights2}
w\bigblp(j,k)^{(1)},(m,n)^{(2)}\bigbrp \\ = w\bigblp(j,k)^{(1)}\bigbrp w\bigblp(m,n)^{(2)}\bigbrp\exp(K\delta_{j,m}\delta_{k,n}/T),
\end{multline}
where the factor $\exp(K\delta_{j,m}\delta_{k,n}/T)$ prevents double-counting of dimer overlap when the bonds $(j,k)$ and $(m,n)$ are identical. \refeq{eq:tp} is then used to obtain the transition probabilities, with $\bm{w}$ a $36$-component vector containing elements $w\bigblp(j,k)^{(1)},(m,n)^{(2)}\bigbrp$. In step 4, the process terminates when both loops close simultaneously, i.e. $k=n=i_{0}$. Otherwise we return to step 2, using $i=k$ and $l=n$.

% Efficiency
Double-loop updates are performed without rejection, but their efficiency is poor at higher temperatures. This is because the probability of simultaneous closure is small, and so updates are unnecessarily long. To reduce this problem, we use double loops only for large \(\lvert K \rvert / T \). We also define a spring potential $V\bigblp(j,k)^{(1)},(m,n)^{(2)}\bigbrp = \frac{1}{2}k\sub{s}\lvert\bm{r}_{k}-\bm{r}_{n}\rvert^{2}$, where $k\sub{s}$ is a spring constant and $\bm{r}_{k}$ denotes the position vector of site $k$. This is imposed by multiplying the equilibrium probabilities of \refeq{eq:weights2} by $\exp(V)$, and favours the selection of sites $k$ and $n$ with smaller separation in step 3. The potential only modifies equilibrium probabilities of configurations during the double-loop construction, and so does not affect detailed balance for the close-packed dimer configurations.

\subsection{Simulation parameters}
\label{simulationparameters}

% Parameters
We performed simulations on systems up to a maximum linear size $L=96$ with PBCs. Following equilibration, data points are typically obtained by averaging over $10^{6}$ MC sweeps, where a sweep is defined such that all lattice bonds are visited once on average. Statistical errors are estimated using a jackknife resampling method. A spring constant $k\sub{s}=2$ is used for double loops.

% Tests
We have checked the MC data by comparing with exact results for the case $L=2$ and with the limits discussed in \refsec{independentreplicas}.

\section{Numerical results}
\label{numericalresults}

% Overview
We have used the worm algorithm to find the phase diagram shown in \reffig{fig:phasediagram}, and to determine the critical properties of each transition. In this section, we present our results for each of the phase boundaries in turn, and also for the nature of the ordered phases at large negative \(J/T\).

\subsection{Synchronized $\longleftrightarrow$ Coulomb}
\label{synchronized-coulomb}

% Introduction
We first focus on the synchronization transition, between the synchronized and Coulomb phases. In particular, we consider the case $J=0$, $K=-1$, and vary the temperature. In the vicinity of the critical point, double-loop updates are not required.

\begin{figure*}
\begin{center}
\includegraphics[width=\textwidth]{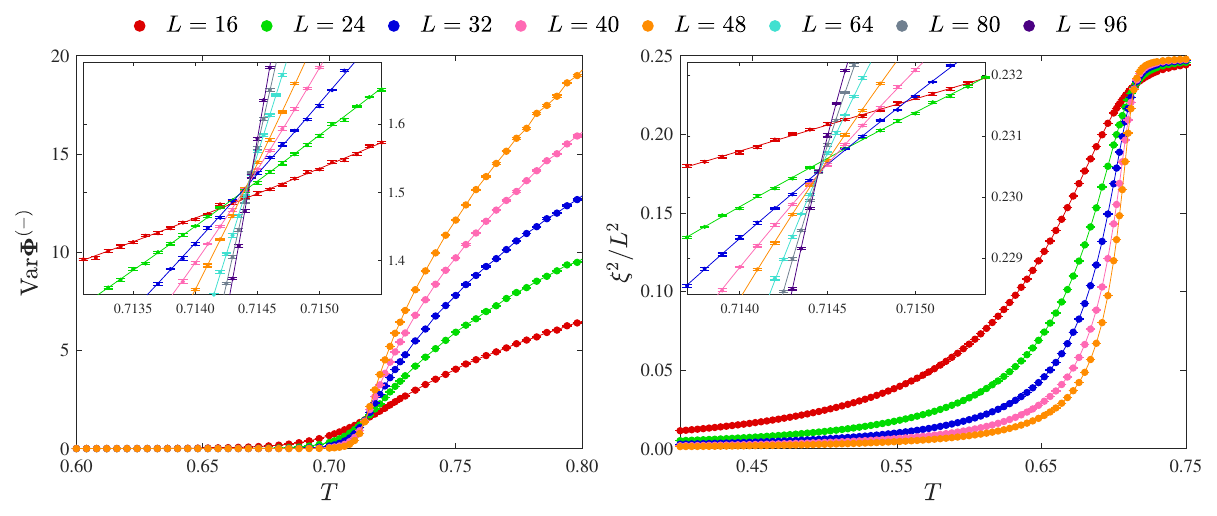}
\caption{Variance of the flux difference \(\Phiv\spr- = \Phiv\spr1 - \Phiv\spr2\) (left panel) and square of the normalized confinement length $\xi^{2}/L^{2}$ (right panel) versus temperature $T$, for the cubic-lattice double dimer model with $J=0$, $K=-1$, and different system sizes $L$. In each case, quadratic fits in the vicinity of the crossing point are shown (insets). Both quantities are small (large) at low (high) temperatures, indicating a phase transition between synchronized and Coulomb phases. The distinct crossing points imply that the transition is continuous.}
\label{fig:cross}
\end{center}
\end{figure*}

% Identify phase transition
Our data for the flux difference variance $\var\Phiv\spr-$ and normalized confinement length $\xi/L$ are shown in \reffig{fig:cross}. Both quantities are small (large) at low (high) temperatures, indicating a phase transition between synchronized and Coulomb phases (see \refsec{coupledreplicas}). In particular, the high-temperature limit $\xi^{2}/L^{2} \simeq 0.25$ is observed, which closely matches the mean-square separation of $(L^2 + 2)/4$ for free monomers hopping on an empty lattice \cite{Chen2009}. This is evidence for deconfined monomers in the Coulomb phase.

\begin{figure}
\begin{center}
\includegraphics[width=\columnwidth]{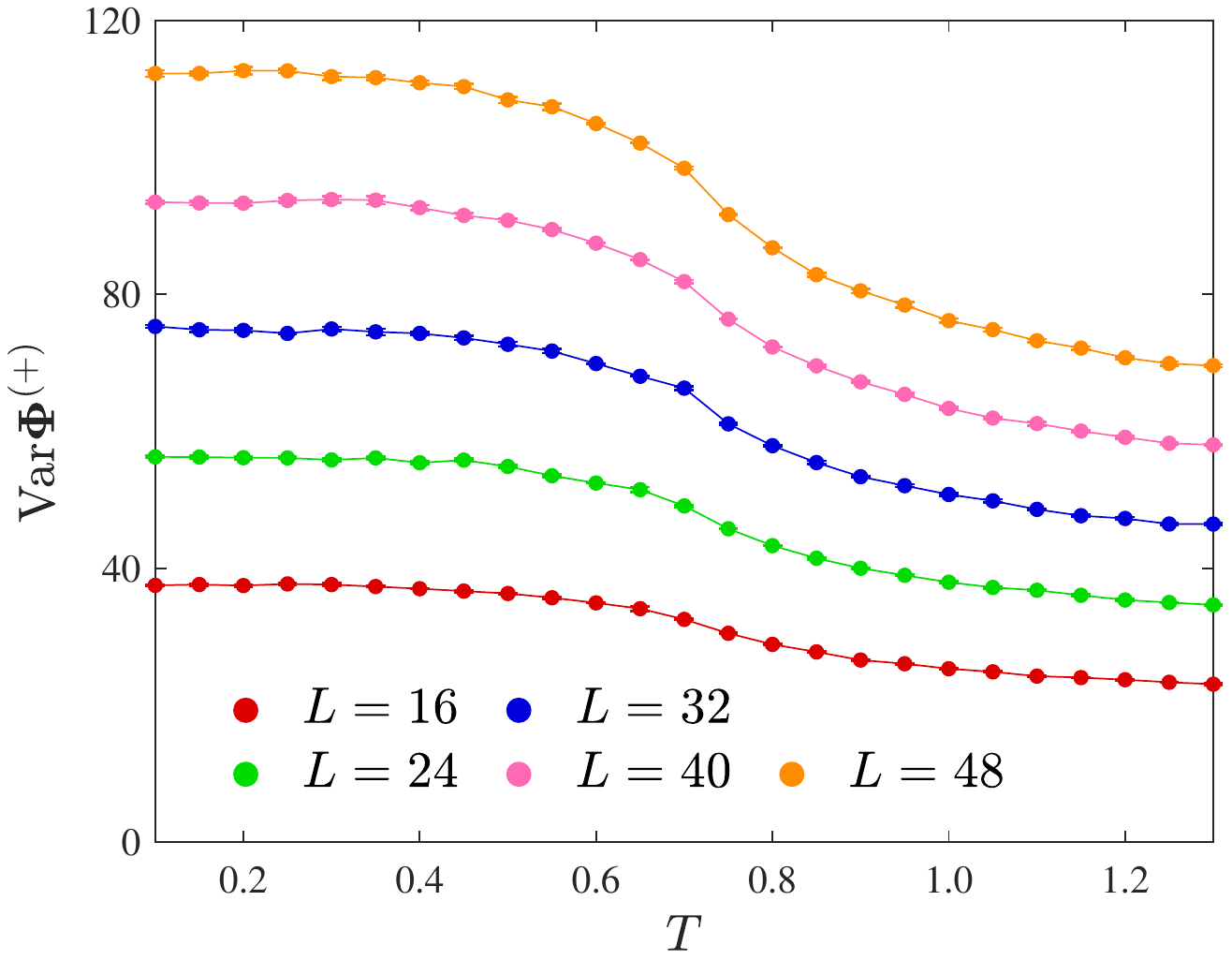}
\caption{Variance of the total flux $\Phiv\spr+ = \Phiv\spr1 + \Phiv\spr2$ versus temperature $T$, for $J=0$, $K=-1$, and different system sizes $L$. At low temperatures the replicas are synchronized, but remain fluctuating, and hence $\var\Phiv\spr+$ is still large (\(\sim L\)).}
\label{fig:totalflux}
\end{center}
\end{figure}

% Total flux
In contrast to $\var\Phiv\spr-$, the variance of the total flux, $\var\Phiv\spr+$, is large in both phases, as shown in \reffig{fig:totalflux}. This confirms that the dimers in each replica remain disordered, even though relative fluctuations between the two replicas are suppressed. In fact, as \(\lvert K\rvert/T\) increases and the replicas become more synchronized, $\var\Phiv\spr+$ becomes larger. In the limit of perfect synchronization, \(K/T = -\infty\), $\Phiv\spr1 = \Phiv\spr2$ and so \(\var\Phiv\spr+ = 4\var\Phiv\spr1\), double the value at \(K = 0\), where $\Phiv\spr{1,2}$ are independent and their variances add.

% Crossing point
In order to classify the phase transition, we use finite-size scaling arguments \cite{Cardy1996}. At the transition of interest, both \(\var\Phiv\spr-\) and \(\xi^{2}/L^{2}\) have zero scaling dimension \cite{Alet2006,Powell2013}, and so for a continuous transition at critical temperature \(T\sub{c}\), obey the scaling forms
\begin{equation}
\label{eq:fdvscaling}
\var\Phiv\spr- \sim f_{\Phi}(L^{1/\nu}t)
\end{equation}
and
\begin{equation}
\label{eq:clscaling}
\xi^{2}/L^{2} \sim f_{\xi}(L^{1/\nu}t)\punc,
\end{equation}
where $t=(T-T\sub{c})/T\sub{c}$ is the reduced temperature, $\nu$ is the correlation-length exponent, and $f_{\Phi}$ and $f_{\xi}$ are universal functions. At the critical temperature $t=0$, \refeqand{eq:fdvscaling}{eq:clscaling} become independent of system size, predicting a distinct crossing point in MC data at $T=T\sub{c}$. This is observed (see \reffig{fig:cross}, insets), indicating that the transition is continuous.

\begin{figure}
\begin{center}
\includegraphics[width=\columnwidth]{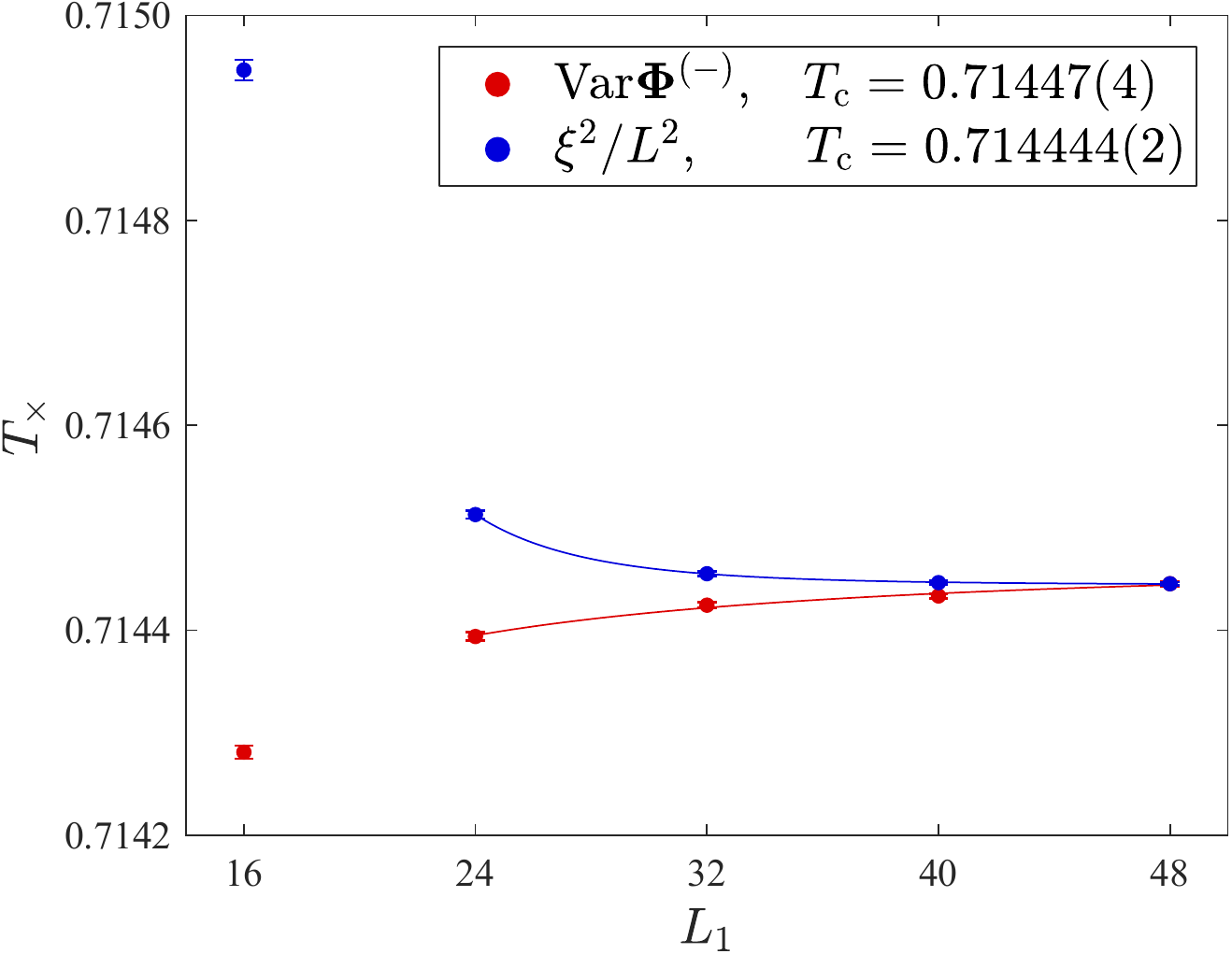}
\caption{Crossing temperature $T_{\times}$ of $\var\Phiv\spr-$ (red) and $\xi^{2}/L^{2}$ (blue), for pairs of system sizes $L_{1},L_{2}$ in the ratio $L_{2}/L_{1}=2$. Solid lines are fits to \refeq{eq:txfit} for $L_{1} \geq 24$, from which consistent values for the critical temperature $T\sub{c}=0.71447(4)$ (flux) and $T\sub{c}=0.714444(2)$ (confinement length) are obtained.}
\label{fig:tx}
\end{center}
\end{figure}

% Critical temperature
In reality, we observe a weak dependence on system size at the critical point, which may be explained by corrections to scaling. Including the leading-order correction, \refeq{eq:fdvscaling} becomes
\begin{equation}
\var\Phiv\spr- \sim f_{\Phi}(L^{1/\nu}t) + uL^{-\lvert y_{u} \rvert}\tilde{f}_{\Phi}(L^{1/\nu}t)\punc,
\end{equation}
where $u$ is a constant, $-\lvert y_{u} \rvert$ is the renormalization group (RG) eigenvalue of the leading irrelevant scaling operator, and $\tilde{f}_{\Phi}$ is a universal function. For two system sizes $L_{1}$ and $L_{2}$, this implies a crossing temperature \(T_\times\) scaling as \cite{Binder1981, Ferrenberg1991}
\begin{equation}
T_{\times}(L_{1}, L_{2}) - T\sub{c} \sim \frac{L_{2}^{-\lvert y_{u} \rvert}-L_{1}^{-\lvert y_{u} \rvert}}{L_{1}^{1/\nu}-L_{2}^{1/\nu}}\punc.
\end{equation}
Fixing the ratio $\rho = L_{2}/L_{1}$ gives
\begin{equation}
\label{eq:txfit}
T_{\times}(L_{1}, \rho L_{1}) - T\sub{c} \sim  L_{1}^{-\lvert y_{u} \rvert-1/\nu}\punc,
\end{equation}
with an identical result applying to the $\xi^{2}/L^{2}$ crossing point. We determine $T\sub{c}$ by fitting to these expressions with $\rho=2$, using quadratic fits in the vicinity of the crossing point to measure \(T_\times\) for each \(L_1\) (insets of \reffig{fig:cross}). From our results, shown in \reffig{fig:tx}, we obtain consistent critical temperatures $T\sub{c}=0.71447(4)$ (flux) and $T\sub{c}=0.714444(2)$ (confinement length).

\begin{figure}
\begin{center}
\includegraphics[width=\columnwidth]{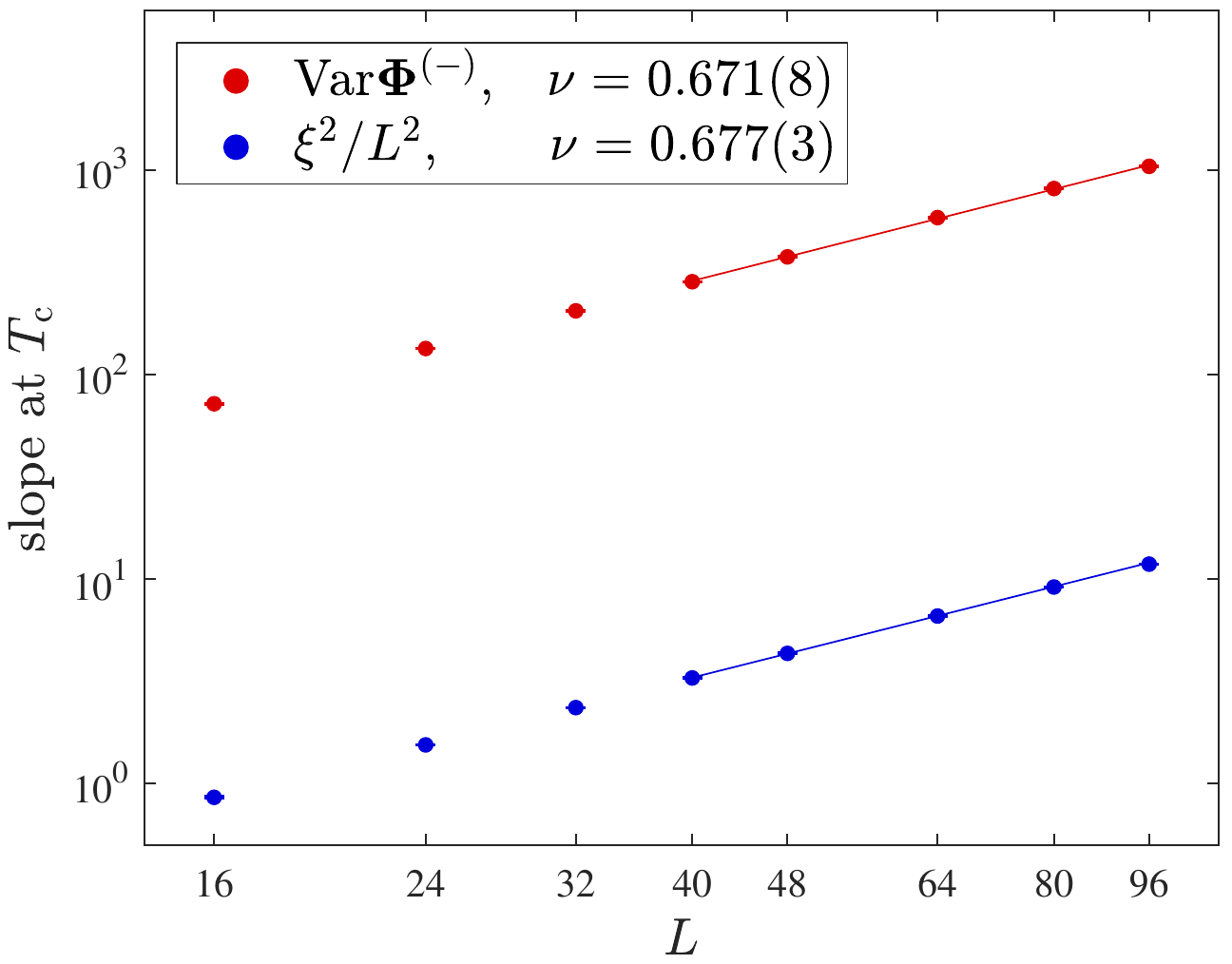}
\caption{Log--log plot of temperature derivative of $\var\Phiv\spr-$ (red) and $\xi^{2}/L^{2}$ (blue), evaluated at the critical temperature $T\sub{c}$, versus system size $L$. Solid lines are fits to \refeq{eq:nufit} for $L \geq 40$, from which consistent values for the correlation length exponent $\nu=0.671(8)$ (flux) and $\nu=0.677(3)$ (confinement length) are obtained.}
\label{fig:nu}
\end{center}
\end{figure}

% Correlation length exponent
In order to determine the correlation length exponent $\nu$, we evaluate the temperature derivative of \refeqand{eq:fdvscaling}{eq:clscaling} at the critical point. For $\var\Phiv\spr-$ this gives
\begin{equation}
\label{eq:nufit}
\left.\frac{\partial}{\partial T}\var\Phiv\spr-\right\rvert_{T=T\sub{c}} \sim L^{1/\nu},
\end{equation}
and one finds the same result for $\xi^{2}/L^{2}$. The system size dependence of the slope at $T\sub{c}$ is extracted from quadratic fits. The results are shown in \reffig{fig:nu}, and a fit to \refeq{eq:nufit} yields consistent estimates $\nu=0.671(8)$ (flux) and $\nu=0.677(3)$ (confinement length). These values are compatible with the 3D XY universality class, for which $\nu\sub{3DXY}=0.6717(1)$ \cite{Campostrini2006}.

\begin{figure}
\begin{center}
\includegraphics[width=\columnwidth]{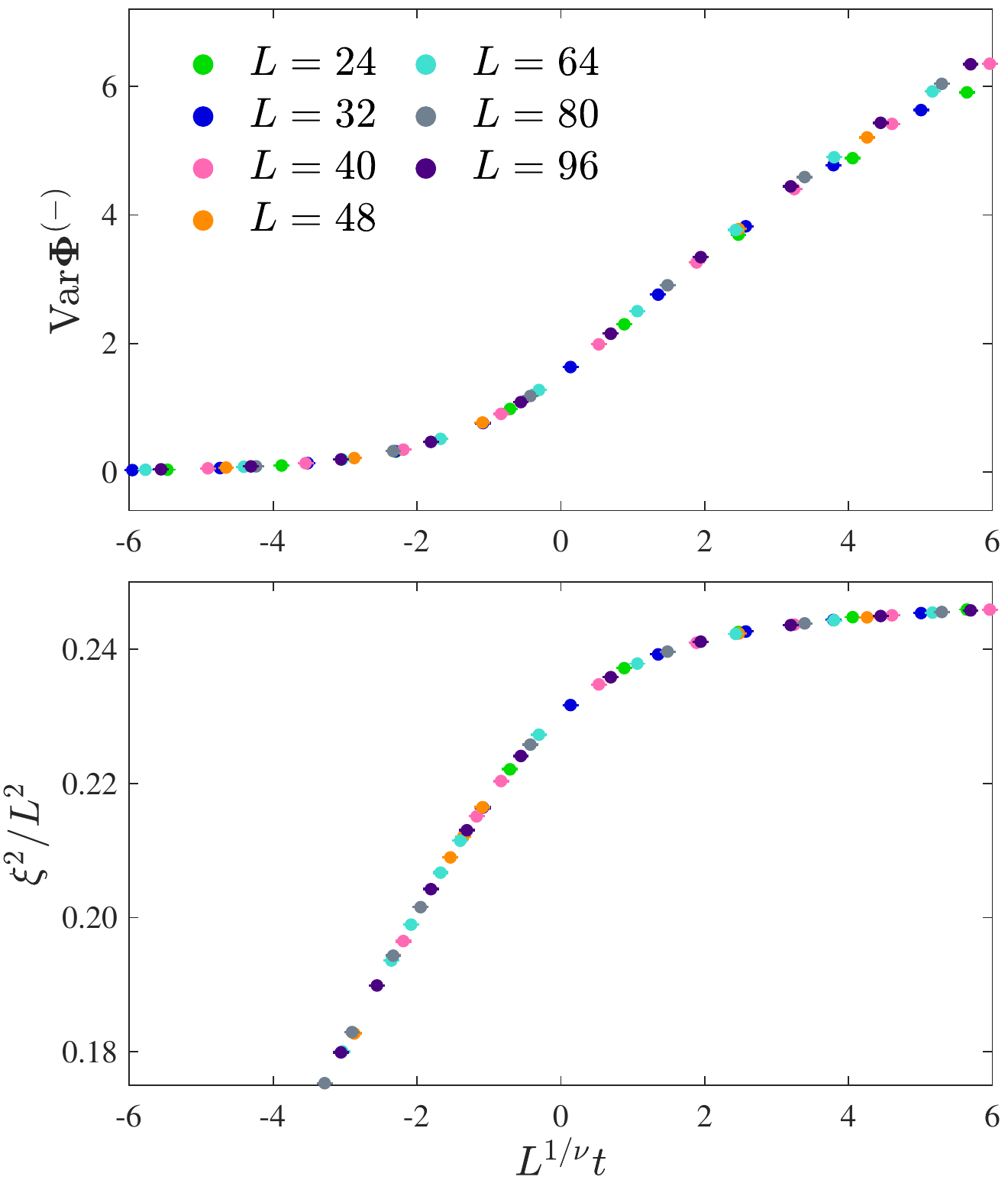}
\caption{Flux difference variance $\var\Phiv\spr-$ (top panel) and squared normalized confinement length $\xi^{2}/L^{2}$ (bottom panel) versus $L^{1/\nu}t$, for $J=0$, $K=-1$, and different system sizes $L$. In each case, we have replotted the data of \reffig{fig:cross} near the critical point, using values $T_{c}=0.714444$ (obtained from the crossing point of the confinement length) and $\nu=\nu\sub{3DXY}=0.6717$. The data collapse is consistent with a synchronization transition in the 3D XY universality class.}
\label{fig:datacollapse}
\end{center}
\end{figure}

% Data collapse
Now equipped with estimates for $T\sub{c}$ and $\nu$, we replot the data of \reffig{fig:cross} against $L^{1/\nu}t$ in \reffig{fig:datacollapse}. Near the critical temperature, a good data collapse is obtained for all but the smallest system size. The curves, which represent the universal functions $f_{\Phi}$ and $f_{\xi}$, are consistent (up to normalization) with those in Fig.~6 of \refcite{Chen2009}.

\begin{figure}
\begin{center}
\includegraphics[width=\columnwidth]{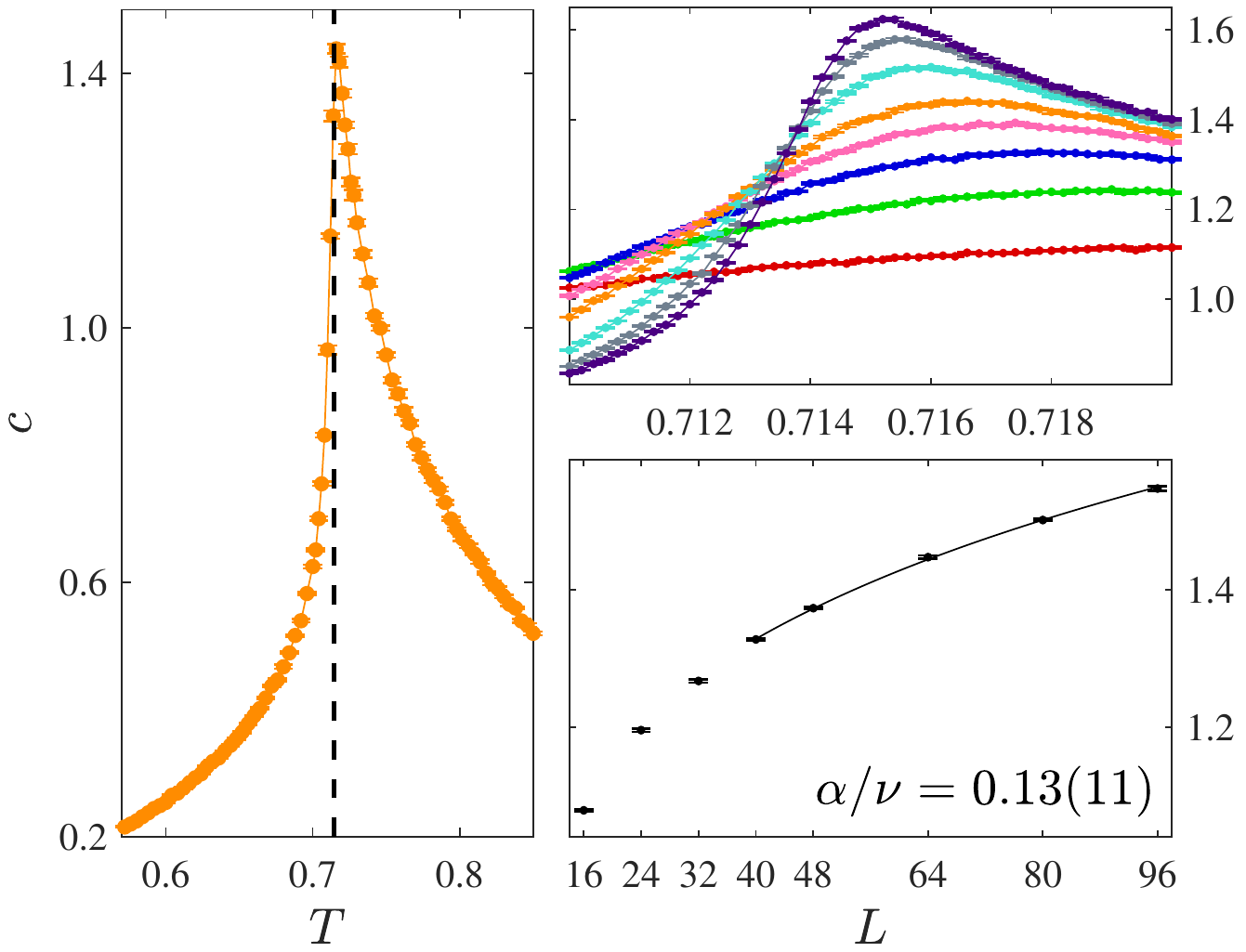}
\caption{Left panel: heat capacity per site $c$ versus temperature $T$ for $J=0$, $K=-1$, and system size $L=48$. A single peak is observed at the critical temperature $T\sub{c}=0.714444$ (dashed vertical line), which grows slowly with system size (top right panel). (Colors indicate different values of $L$ as in \reffig{fig:cross}.) Bottom right panel: System size dependence at the critical temperature $T\sub{c}=0.714444$. The solid line is a fit to \refeq{eq:cfit} for $L \geq 40$, from which a value $\alpha/\nu=0.13(11)$ is obtained.}
\label{fig:heatcap}
\end{center}
\end{figure}

% Specific heat exponent
As shown in \reffig{fig:heatcap} (left panel), a single peak in the heat capacity per site $c$ is observed at the transition temperature, indicating a single phase transition between the synchronized and Coulomb phases. To measure the specific heat exponent $\alpha$, we consider its scaling at the critical point,
\begin{equation}
\label{eq:cfit}
c = c_{0} + AL^{\alpha/\nu},
\end{equation}
where $c_{0}$ represents the regular part, and $A$ is a constant. A fit to this form in \reffig{fig:heatcap} (bottom right panel) yields $\alpha/\nu=0.13(11)$, and using $\nu=0.677(3)$ (flux) gives a rough estimate $\alpha=0.09(7)$. In the 3D XY universality class, the corresponding value is $\alpha\sub{3DXY}=-0.0151(3)$ \cite{Campostrini2006}. Our results satisfy hyperscaling $\alpha = 2-d\nu$, where $d=3$ is the spatial dimension.

\begin{figure}
\begin{center}
\includegraphics[width=\columnwidth]{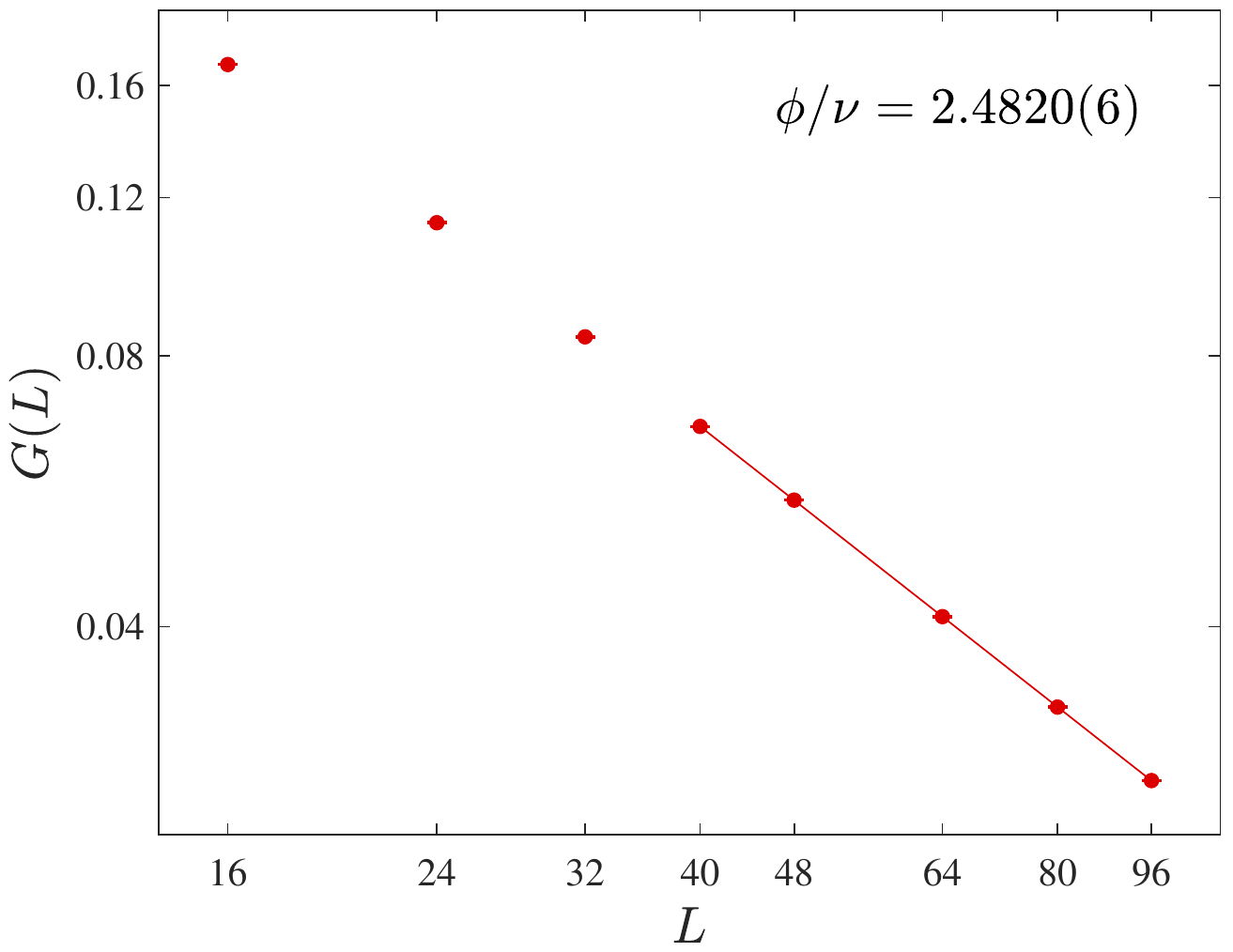}
\caption{Log--log plot of $G(L)$, the normalized value of the monomer distribution function $G\sub{m}$, evaluated at the critical temperature $T\sub{c}=0.714444$, versus system size $L$. The solid line is a fit to \refeq{eq:yzfit} for $L \geq 40$, from which a value $\phi/\nu=2.4820(6)$ is obtained.}
\label{fig:yz}
\end{center}
\end{figure}

% Crossover exponent
We next measure the crossover exponent $\phi$, which can be found by considering the monomer distribution function $G\sub{m}$ \cite{Powell2013}. Each MC simulation can only construct $G\sub{m}$ up to an arbitrary multiplicative constant, so we define the ratio
\begin{equation}
G(L) = \frac{G\sub{m}(\bm{R}_{\text{max}};L)}{G\sub{m}(\bm{R}_{\text{min}};L)},
\end{equation}
where $\lvert \bm{R}_{\text{max}} \rvert \sim L$, $\lvert \bm{R}_{\text{min}} \rvert = 1$, and the system size dependence of $G\sub{m}$ has been shown explicitly. At the critical point, this has scaling form \cite{Sreejith2014}
\begin{equation}
\label{eq:yzfit}
G(L) \sim L^{-2\left(d-\frac{\phi}{\nu}\right)},
\end{equation}
for sufficiently large systems. A fit to this form in \reffig{fig:yz} yields $\phi/\nu=2.4820(6)$, and using $\nu=0.677(3)$ (confinement length) gives $\phi=1.680(8)$. This value is compatible with the 3D XY universality class, for which $\phi_{\textrm{3DXY}}=d\nu_{\textrm{3DXY}}-\beta_{\textrm{3DXY}}=1.6665(3)$, using the exponents reported in \refcite{Campostrini2006}.

% J \= 0
Finally, we consider the same phase boundary, between the synchronized and Coulomb phases, at points where $J \neq 0$. The critical point \((K/T)\sub{c}\) for each \(J/T\), plotted in \reffig{fig:phasediagram}, has been obtained from the crossing point of $\var\Phiv\spr-$ for system sizes $L=16$ and $L=24$. We expect that the universality class is the same for each point along the boundary, and have confirmed this for the points $J/T = -0.16$ and $J/T=-0.24$.

\subsection{Columnar \& (Anti)synchronized $\longleftrightarrow$ Coulomb}
\label{columnar&synchronized-coulomb}

% Introduction
As discussed in \refsec{independentreplicas}, independent replicas ($K=0$) exhibit a continuous transition between the columnar and Coulomb phases \cite{Alet2006}. We now consider columnar ordering of coupled replicas ($K\neq 0$), i.e., the transition between the columnar \& (anti)synchronized and Coulomb phases. Our results indicate that columnar ordering is driven first-order when replicas are coupled. (Certain other additional interactions have previously been shown to have this effect \cite{Charrier2010, Papanikolaou2010}.)

\begin{figure}
\begin{center}
\includegraphics[width=\columnwidth]{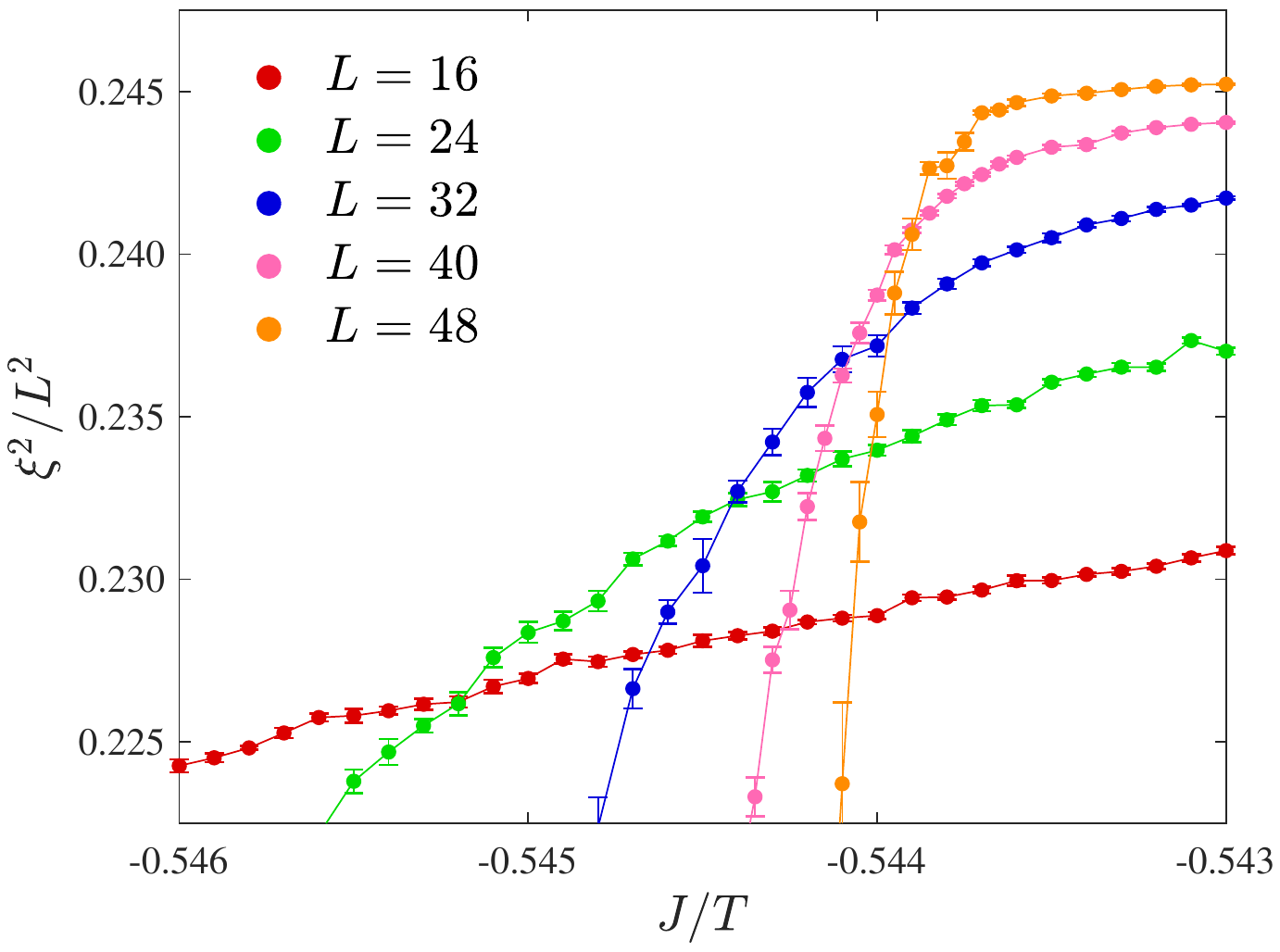}
\caption{Squared normalized confinement length $\xi^{2}/L^{2}$ versus $J/T$, for $K/T=-0.2$, $J/T \simeq (J/T)\sub{c}$, and different system sizes $L$. A confinement transition between the columnar \& synchronized and Coulomb phases is not accompanied by a distinct crossing point, and is thus not continuous.}
\label{fig:nocross}
\end{center}
\end{figure}

% No crossing point
According to \refeq{eq:clscaling}, a continuous (confinement) transition is characterized by a crossing point in $\xi^{2}/L^{2}$, at the critical temperature. We plot this quantity in \reffig{fig:nocross}, in the vicinity of a transition between the columnar \& synchronized and Coulomb phases. A distinct crossing point is not observed [cf.\ \reffig{fig:cross} (insets)], and hence the transition is not continuous. Similar behaviour is obtained for $\var\Phiv\spr\pm$.

\begin{figure}
\begin{center}
\includegraphics[width=\columnwidth]{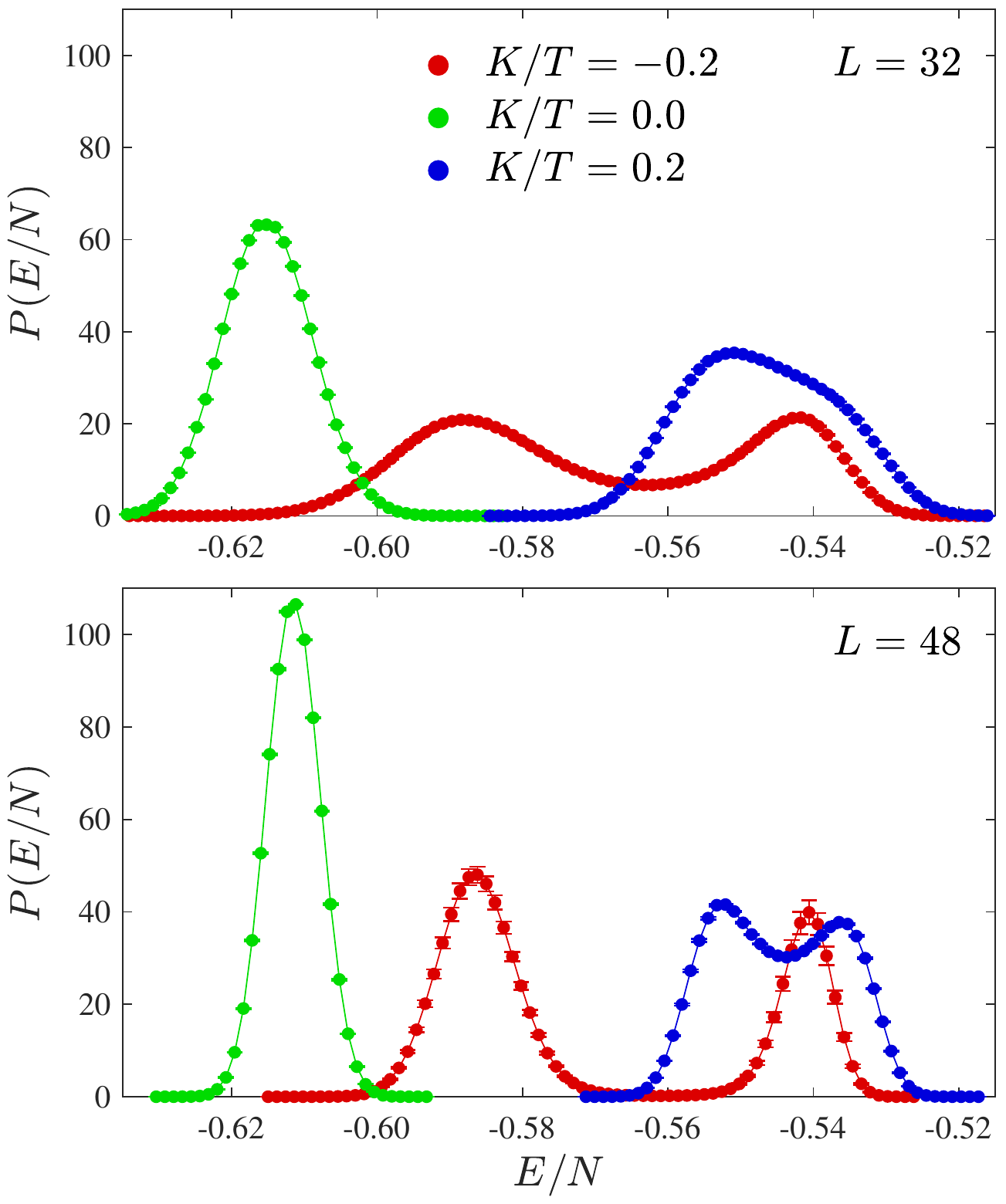}
\caption{Histograms of energy per site, \(E/N\), for different values of $K/T$ and with \(J/T\) close to the columnar-ordering transition for each. The system sizes shown are \(L=32\) (top panel) and \(L=48\) (bottom panel). A single peak at $K/T=0$ is consistent with the well-known continuous transition between columnar and Coulomb phases. The distributions for $K/T \neq 0$ are mixtures of two approximately normal distributions, and become more clearly bimodal for larger \(L\), indicating first-order transitions between columnar \& (anti)synchronized and Coulomb phases.}
\label{fig:energyhistogram}
\end{center}
\end{figure}

% Bimodal energy histogram
Instead, the transition must be first-order. One thus expects a bimodal energy histogram in the vicinity of the critical point, which can be seen in \reffig{fig:energyhistogram} (red). The same behavior is also obtained for transitions between the columnar \& antisynchronized and Coulomb phases (blue). In contrast, a single peak is observed for columnar ordering of independent replicas (green), as expected for a continuous transition.

% Phase diagram
Six points along this first-order phase boundary are included in the phase diagram of \reffig{fig:phasediagram}. These have been located by identifying peaks in the heat capacity per site, using system size $L=32$.

\subsection{Columnar \& Synchronized $\longleftrightarrow$ Synchronized}
\label{columnar&synchronized-synchronized}

% Expectations
We next consider the transition between the columnar \& synchronized and synchronized phases. In the limit $K/T \rightarrow -\infty$, this phase boundary corresponds to columnar ordering of a single dimer model with $J_{\textrm{eff}}=2J$ (see \refsec{coupledreplicas}). This is known to be an (apparently) continuous transition in the tricritical universality class, and we expect the whole phase boundary to share the same critical properties as this point.

% Phase diagram
Since the flux difference variance $\var\Phiv\spr-$ and confinement length $\xi$ are small in both (synchronized) phases, we locate the phase boundary using crossing points in the total flux variance $\var\Phiv\spr+$, for system sizes $L=16$ and $L=24$. We have analyzed the points $K/T = -2.0$ and $K/T = -1.2$ in greater detail (not shown), and verified the expected critical properties.

\subsection{Columnar \& (Anti)synchronized phases}
\label{columnar&synchronized-columnar&antisynchronized}

% Introduction
Finally, we consider the different possible columnar-ordered phases at negative \(J/T\) and both signs of \(K/T\). To classify these, it is convenient to use the covariance of the replica magnetizations $\sigma_{12}=\langle\Mv\spr{1}\cdot \Mv\spr{2}\rangle$, which, deep within the columnar-ordered region, indicates the relative orientations of the magnetizations.

% Columnar and columnar & synchronized
In the columnar phase at \(K=0\), the two replicas are independent, and so $\sigma_{12}=\langle\Mv\spr{1}\rangle \cdot \langle\Mv\spr{2}\rangle=0$, since the mean magnetization vanishes by symmetry. Deep within the columnar \& synchronized phase for \(K<0\), the 6 ground states with $\Mv\spr{1} = \Mv\spr{2}= \pm \deltav_\mu$ [see \reffig{fig:groundstates}(a)] dominate, giving $\sigma_{12}=1$.

% Columnar & antisynchronized
For positive \(K\), there are two sets of configurations that minimize the energy: 6 where the magnetizations are antiparallel, $\Mv\spr{1} = -\Mv\spr{2}= \pm \deltav_\mu$ [see \reffig{fig:groundstates}(b)], and \(6\times 4 = 24\) where they are perpendicular, $\Mv\spr{1} \cdot \Mv\spr{2}= 0$ [see \reffig{fig:groundstates}(c)]. Because the degeneracy between the two sets is accidental (i.e., not required by symmetry), it is liable to be resolved by order by disorder (ObD). There are, \emph{a priori}, three possibilities: ObD favoring antiparallel magnetizations; ObD favoring perpendicular magnetizations; and no ObD, leaving all orientations equally likely. For large negative \(J/T\), where columnar order is well established and so \(\langle\lvert\Mv\spr\alpha\rvert\rangle \simeq 1\), these give limiting values of
\beq[eq:covariance]
\sigma_{12}=
% Use aligned rather than cases for right alignment
\left\{\begin{aligned}
-1 & \qquad\text{ObD, antiparallel}\\
0 & \qquad\text{ObD, perpendicular}\\
-0.2 & \qquad\text{no ObD.}
\end{aligned}\right.
\eeq

% Results
\begin{figure}
\begin{center}
\includegraphics[width=\columnwidth]{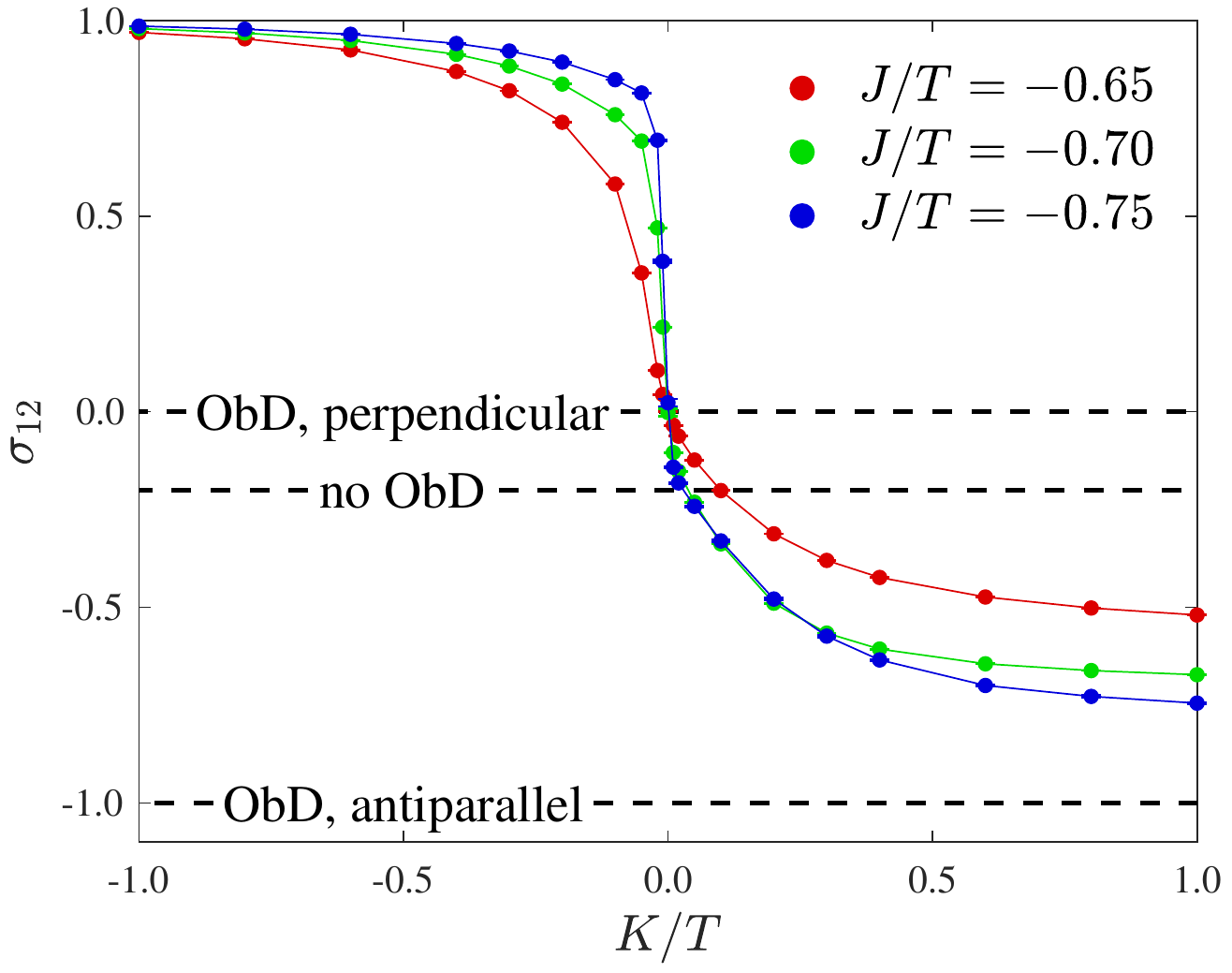}
\caption{Covariance of the replica magnetizations $\sigma_{12}=\langle\Mv\spr{1}\cdot \Mv\spr{2}\rangle$ versus $K/T$, for different values of $J/T<0$ and system size $L=8$. As explained in \refsec{columnar&synchronized-columnar&antisynchronized}, \(\sigma_{12}\) vanishes at \(K=0\) and approaches \(+1\) in the columnar \& synchronized phase for \(K/T < 0\). Deep within the columnar \& antisynchronized phase, \(K/T > 0\), we expect \(\sigma_{12}\) to approach one of the values in \refeq{eq:covariance}, shown with dashed lines, depending on the result of order-by-disorder effects. The evidence indicates that antiparallel magnetizations \(\Mv\spr{1} = -\Mv\spr{2}\) are preferred. (Accessible values of \(L\) and \(\lvert J\rvert /T\) are limited by loss of ergodicity deep within the ordered phase.)}
\label{fig:orderbydisorder}
\end{center}
\end{figure}

MC results for $\sigma_{12}$ are shown in \reffig{fig:orderbydisorder}. The expected behaviour is obtained in the columnar phase ($\sigma_{12}=0$ when $K/T=0$), and the columnar \& synchronized phase ($\sigma_{12} \rightarrow 1$ for $K/T<0$). In the columnar \& antisynchronized phase, the data appear to converge towards $\sigma_{12}=-1$ as $J/T$ becomes more negative. This indicates that ObD selects an arrangement with antiparallel magnetizations, in agreement with consideration of the elementary fluctuations, as in \refsec{coupledreplicas}.

\section{Bethe lattice}
\label{bethelattice}

To gain further insight into the synchronization transition, we consider the double dimer model on the Bethe lattice, which, we will show, can be solved exactly. This provides an approximation to the model on the cubic lattice that is in the spirit of mean-field theory. In particular, we expect it to reproduce the qualitative behavior correctly, with a critical temperature that approximates the true value, but to fail to predict the critical properties.

\begin{figure}
\begin{center}
\includegraphics[width=\columnwidth]{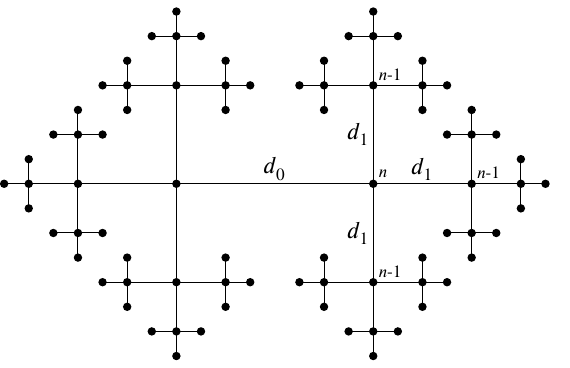}
\caption{Cayley tree with coordination number $q=4$. The `root' bond, labeled \(d_0\), has depth \(n\) and is connected at each of its vertices to \(q-1\) `subbranches', with roots labeled \(d_1\), of depth \(n-1\).}
\label{fig:cayleytree}
\end{center}
\end{figure}

% Cayley tree and Bethe lattice
We first consider a `Cayley tree', illustrated in \reffig{fig:cayleytree}, a graph where each vertex has \(q\) neighbors, except for those at the boundaries, and where there are no closed loops. To avoid contributions from the boundaries, which in the thermodynamic limit constitute a finite fraction of the vertices, we define the Bethe lattice as the part of the Cayley tree that is far away from all boundaries. A dimer model on the Bethe lattice has dimers occupying the bonds of the lattice (i.e., the edges of the graph).

% Exact solutions
Statistical mechanics problems with nearest-neighbour interactions are often exactly solvable on the Bethe lattice, since the absence of circuits allows one to formulate a recurrence relation for the partition function. This method has been used for the Ising model \cite{Baxter1982}, whilst a similar calculation has been performed for spin ice on the Husimi tree \cite{Jaubert2008, Jaubert2009, Powell2013}. Here we apply it to the synchronization transition on the Bethe lattice.

\subsection{Noninteracting dimers}
\label{noninteractingdimers}

% Quantity of interest
To illustrate the method, we begin with a simpler calculation. Consider a single close-packed dimer model, with no interactions, on the Cayley tree. In this case the partition function is simply
\begin{equation}
Z = \sum_{c\in\goC_0}1\punc.
\end{equation}
A quantity of interest is the mean dimer occupation number for the central bond, or root, of the Cayley tree, given by
\begin{equation}
\label{eq:rootcf}
\langle d_{0}\rangle = \frac{1}{Z}\sum_{c\in\goC_0}d_{0}\punc.
\end{equation}

% Z_n
To begin, the partition function is written as
\begin{align}
\label{eq:e.g.}
Z &= \sum_{d_{0}}\left[Z_{n}(d_{0})\right]^2 \\
\label{eq:fullpf}&=\left[Z_{n}(0)\right]^{2}+\left[Z_{n}(1)\right]^{2}\punc.
\end{align}
In the first line, the quantity $Z_{n}(d_{0})$ is the `partial partition function' of the left, or equivalently right, branch of the Cayley tree, when the root dimer occupation number is fixed to $d_{0}$. The index $n$ enumerates the branch depth. The same logic may be applied to \refeq{eq:rootcf}, and results in
\begin{equation}
\label{eq:rootcf2}
\langle d_{0}\rangle =  \frac{\left[Z_{n}(1)\right]^{2}}{\left[Z_{n}(0)\right]^{2}+\left[Z_{n}(1)\right]^{2}}\punc.
\end{equation}
A branch with root $d_{0}$ and depth $n$ consists of $(q-1)$ `subbranches', rooted at $d_{1}$ and with depth $n-1$ (see \reffig{fig:cayleytree}). This observation allows the construction of recurrence relations which connect the partial partition functions of branches of depth $n$ and $n-1$. By allowing for all consistent configurations of the subbranches, while applying (at the roots) the constraint that each site should be covered by exactly one dimer, one finds
\begin{align}
\label{eq:rz0}
Z_{n}(0) &= (q-1)Z_{n-1}(1)\left[Z_{n-1}(0)\right]^{q-2}\\
\label{eq:rz1}
Z_{n}(1) &= \left[Z_{n-1}(0)\right]^{q-1}\punc.
\end{align}

% x
It is convenient to introduce the variable
\begin{equation}
\label{eq:xn}
x_{n} = \frac{Z_{n}(0)}{Z_{n}(1)}\punc,
\end{equation}
for which \refeqand{eq:rz0}{eq:rz1} imply
\begin{equation}
\label{eq:rec}
x_{n} = \frac{q-1}{x_{n-1}}\punc.
\end{equation}
Next we consider only sites on the Bethe lattice, deep within the Cayley tree, by taking the thermodynamic limit $n\rightarrow\infty$. Here, the solution is a fixed point satisfying $x_{n}=x_{n-1}=x$, so that \refeq{eq:rootcf2} may be re-written
\begin{equation}
\label{eq:fix}
\langle d_{0} \rangle = \frac{1}{1+x^{2}}\punc,
\end{equation}
whilst \refeq{eq:rec} becomes
\begin{equation}
x = \frac{q-1}{x}\punc.
\end{equation}
This has solution
\begin{equation}
\label{eq:x}
x = \sqrt{q-1}\punc,
\end{equation}
and substitution into \refeq{eq:fix} yields
\begin{equation}
\label{eq:singlerootcf3}
\langle d_{0} \rangle = \frac{1}{q}\punc.
\end{equation}
This is given, as expected, by the ratio of the number of dimers to the number of bonds.

% Convergence
In reality, the recurrence relation of \refeq{eq:rec} does not converge towards its fixed point in the thermodynamic limit, but instead oscillates indefinitely. To perform a more rigorous treatment, one can permit monomers with a small nonzero fugacity $z$, modifying \refeq{eq:rec} to
\begin{equation}
x_{n} = \frac{q-1}{x_{n-1}}+z\punc.
\end{equation}
This recurrence relation does converge in the thermodynamic limit, and \refeq{eq:singlerootcf3} is easily retrieved by subsequently taking the limit $z\rightarrow0$.

% Monomer on root
Using the same approach, one can also calculate the response to monomer insertion, which, as discussed in \refsec{phasediagram}, allows one to distinguish confined and deconfined phases. The distribution function \(G\sub{m}\), defined in \refeq{EqGm}, involves a pair of monomers and cannot easily be calculated using the recurrence relation. Instead, we consider the corresponding quantity for a single monomer,
\beq[eq:gminf]
\Psi\sub{m} = \frac{Z\sub{m}}{Z}\punc,
\eeq
where $Z\sub{m} = \sum_{c\in\goC(\rv_+)} \ee^{-E/T}$ is the partition function with a monomer inserted at \(\rv_+\). This takes the form of an expectation value (specifically, of a monomer insertion operator \cite{Sreejith2019}) and we therefore refer to it as the `monomer expectation value'. While it vanishes due to the requirement of charge neutrality when periodic boundary conditions are applied, it can be nonzero with open boundary conditions, including on the Bethe lattice.

Suppose \(\rv_+\) is taken as the left side of the root $d_{0}$. Then the left (right) branch of the Cayley tree `sees' an occupied (unoccupied) root, and the system has partition function $Z\sub{m}=Z_{n}(1)Z_{n}(0)$. The partition function without monomers $Z$ is again given by \refeq{eq:fullpf}. From the definition of \refeq{eq:xn}, and its solution in \refeq{eq:x}, one obtains
\begin{equation}
\Psi\sub{m}=\frac{\sqrt{q-1}}{q}.
\end{equation}
The result is nonzero, indicating that an isolated monomer can occur with finite free-energy cost \(\Delta F\sub{m} = -T \ln \Psi\sub{m}\). Monomers are therefore deconfined, as expected in the noninteracting dimer model.

\subsection{Synchronization transition}
\label{synchronizationtransition}

% Quantity of interest
Now consider the double dimer model of \refeq{eq:configurationenergy} on the Cayley tree. Since parallel pairs of dimers cannot be defined, we set $J=0$, leaving configuration energies
\begin{equation}
E = K\sum_{l}d_{l}^{(1)}d_{l}^{(2)}\punc,
\end{equation}
and a partition function given by \refeq{eq:partitionfunction}. The quantity of interest is the mean energy per site deep within the interior of the tree, which we take as its value on the root bond,
\begin{equation}
\label{eq:energypersite}
\frac{\langle E\rangle}{N}=\frac{q}{2}K\langle d_{0}^{(1)}d_{0}^{(2)}\rangle\punc,
\end{equation}
assuming translational symmetry (at least on average). We therefore require the correlation function
\begin{equation}
\label{eq:rootcf3}
\langle d_{0}^{(1)}d_{0}^{(2)}\rangle = \frac{1}{Z}\sum_{\substack{c\spr{1}\in\goC_0\\c\spr{2}\in\goC_0}}d_{0}^{(1)}d_{0}^{(2)}\ee^{-E/T}
\end{equation}
on the same bond. This may be calculated in analogy with \refsec{noninteractingdimers}, although the algebra is more involved.

% Z_n
\begin{widetext}
The partition function is written as
\begin{align}
Z &= \sum_{d_{0}^{(1)},d_{0}^{(2)}}\ee^{-kd_{0}^{(1)}d_{0}^{(2)}}\left[Z_{n}\left(d_{0}^{(1)}, d_{0}^{(2)}\right)\right]^2 \\
\label{eq:fullpf2} &=\left[Z_{n}(0,0)\right]^{2}+\left[Z_{n}(1,0)\right]^{2}+\left[Z_{n}(0,1)\right]^{2}+\ee^{-k}\left[Z_{n}(1,1)\right]^{2}\punc,
\end{align}
where the reduced coupling $k=K/T$ has been introduced for convenience. In the first line, the quantity $Z_{n}\left(d_{0}^{(1)}, d_{0}^{(2)}\right)$ is the `partial partition function' of the left, or equivalently right, branch of the Cayley tree, when the root dimer occupation numbers are fixed to $d_{0}^{(1)}$ and $d_{0}^{(2)}$. Similarly, \refeq{eq:rootcf3} may be written
\begin{equation}
\label{eq:rootcf4}
\langle d_{0}^{(1)}d_{0}^{(2)}\rangle =  \frac{\ee^{-k}\left[Z_{n}(1,1)\right]^{2}}{\left[Z_{n}(0,0)\right]^{2}+\left[Z_{n}(1,0)\right]^{2}+\left[Z_{n}(0,1)\right]^{2}+\ee^{-k}\left[Z_{n}(1,1)\right]^{2}}\punc.
\end{equation}
In order to construct recurrence relations, one must again allow for all possible configurations of the subbranches, while applying (at the roots) the constraint that each site should be covered by exactly one dimer in each replica. The results are
\begin{gather}
\label{eq:rz00}
Z_{n}(0,0) = (q-1)\ee^{-k}Z_{n-1}(1,1)\left[Z_{n-1}(0,0)\right]^{q-2} + (q-1)(q-2)Z_{n-1}(1,0)Z_{n-1}(0,1)\left[Z_{n-1}(0,0)\right]^{q-3}\\
Z_{n}(1,0) = (q-1)Z_{n-1}(0,1)\left[Z_{n-1}(0,0)\right]^{q-2}\\
Z_{n}(0,1) = (q-1)Z_{n-1}(1,0)\left[Z_{n-1}(0,0)\right]^{q-2}\\
\label{eq:rz11}
Z_{n}(1,1) = \left[Z_{n-1}(0,0)\right]^{q-1}\punc.
\end{gather}
\end{widetext}

% uvw
Next we define the variables
\begin{equation}
\label{eq:unvnwn}
\begin{pmatrix} u_{n} \\ v_{n} \\ w_{n} \end{pmatrix} = \frac{1}{Z_{n}(1,1)} \begin{pmatrix} Z_{n}(0,0) \\ Z_{n}(1,0) \\ Z_{n}(0,1) \end{pmatrix}
\end{equation}
and take the thermodynamic limit. The solutions are again fixed points, and \refeq{eq:rootcf4} may be rewritten
\begin{equation}
\label{eq:rootcf5}
\langle d_{0}^{(1)}d_{0}^{(2)}\rangle =  \frac{\ee^{-k}}{u^{2}+v^{2}+w^{2}+\ee^{-k}}\punc,
\end{equation}
whilst \refeqs{eq:rz00}{eq:rz11} translate into a system of coupled, nonlinear equations given by
\begin{gather}
u = \frac{q-1}{u}\left[\ee^{-k}+(q-2)\frac{vw}{u}\right]\\
v = (q-1)\frac{w}{u}\\
w = (q-1)\frac{v}{u}\punc.
\end{gather}

The solutions to this system depend on the value of the reduced coupling \(k\). For \(k \le k\sub{c} = -\log(q-1)\), there is a single solution
\beq[eq:uvw1]
\begin{pmatrix} u \\ v \\ w \end{pmatrix}=\begin{pmatrix} \sqrt{(q-1)\ee^{-k}} \\ 0 \\ 0 \end{pmatrix}\punc,
\eeq
whereas, for \(k > k\sub{c}\), we find additionally
\beq[eq:uvw2]
\begin{pmatrix} u \\ v \\ w \end{pmatrix}=
\begin{pmatrix} q-1 \\ \sqrt{\frac{q-1}{q-2}(q-1-\ee^{-k})} \\ \sqrt{\frac{q-1}{q-2}(q-1-\ee^{-k})} \end{pmatrix}\punc,
\eeq
which, as we have confirmed by a linear stability analysis, is the only stable solution. (Note that the critical value \(k\sub{c}\) is negative -- as expected, the transition occurs for attractive coupling \(K < 0\).)

Substitution of this result into \refeq{eq:rootcf5}, and then into \refeq{eq:energypersite}, yields the final result for the mean energy per site of
\begin{equation}
\label{eq:energypersite2}
\frac{\langle E\rangle}{N}=
\begin{cases}
\dfrac{K}{2}\dfrac{(q-2)\ee^{-k}}{(q-1)^{2}-\ee^{-k}} & \text{for \(k\ge k\sub{c}\)} \\
\dfrac{K}{2} & \text{\phantom{for} \(k \le k\sub{c}\).}
\end{cases}
\end{equation}

% Convergence
In line with $x_{n}$ in \refsec{noninteractingdimers}, the recurrence relations for $u_{n}$, $v_{n}$ and $w_{n}$ derivable from \refeqs{eq:rz00}{eq:rz11} may oscillate in the thermodynamic limit. Again, convergence is achieved by allowing monomers with fugacity $z$ and taking the limit $z \rightarrow 0$.

\begin{figure}
\begin{center}
\includegraphics[width=\columnwidth]{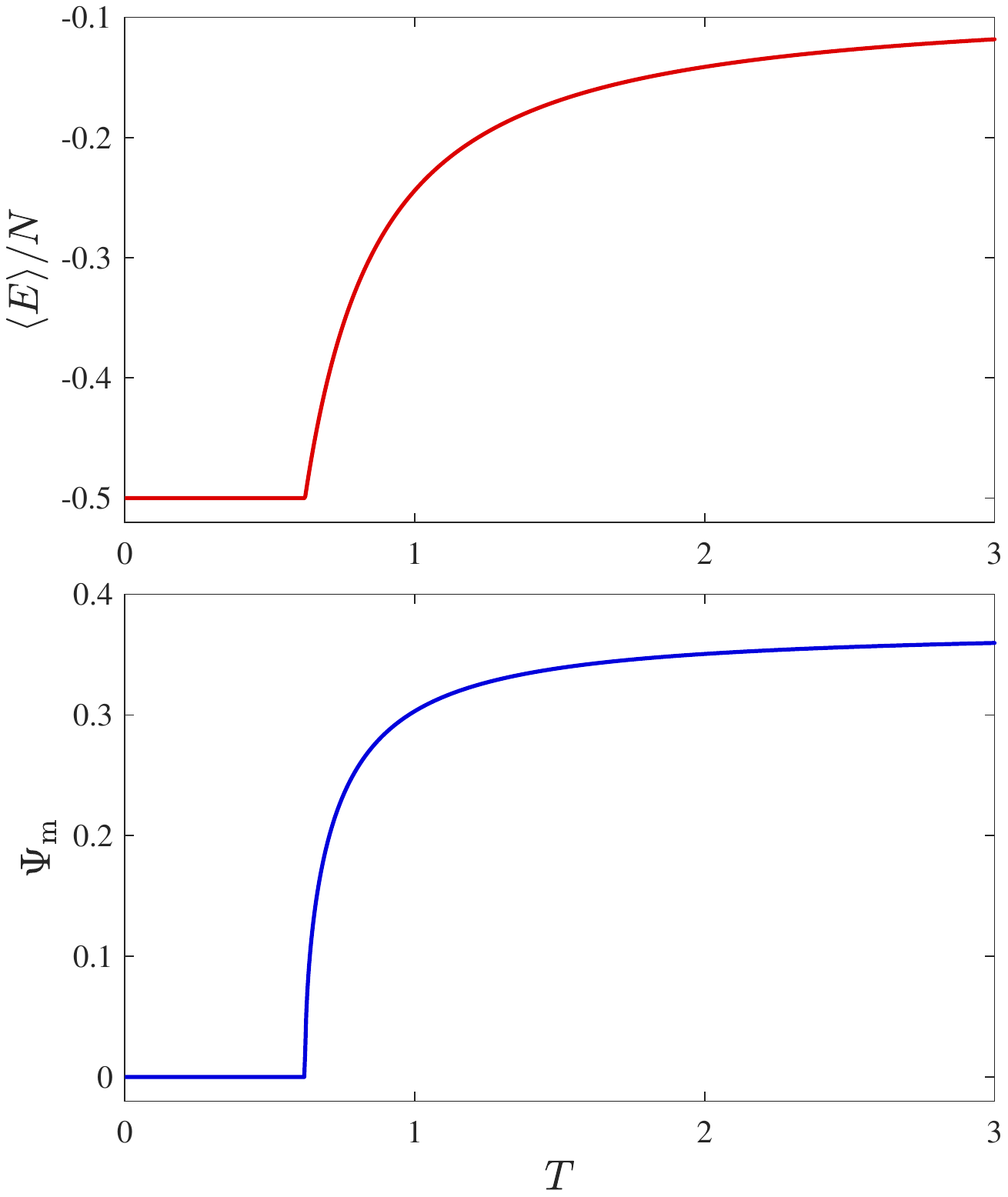}
\caption{Mean energy per site $\langle E\rangle/N$ (top panel) and monomer expectation value $\Psi\sub{m}$ (bottom panel), versus temperature $T$ for the double dimer model on a Bethe lattice with the same coordination number as the cubic lattice, $q=6$. In this case, there are no interactions within each replica (i.e., $J=0$), and we set $K=-1$. A second-order phase transition at $T\sub{c}=1/\log(5) \simeq 0.62$ separates a low-temperature (perfectly) synchronized phase, in which monomers are confined (\(\Psi\sub{m}=0\)), from a high-temperature unsynchronized phase, in which monomers are deconfined (\(\Psi\sub{m}>0\)).}
\label{fig:bethelattice}
\end{center}
\end{figure}

% Discussion
In \reffig{fig:bethelattice} (top panel) we show the temperature dependence of the mean energy per site for a Bethe lattice with the same coordination number as the cubic lattice, $q=6$, and with $K=-1$. There is a second-order phase transition at $T\sub{c}=1/\log(5)\simeq 0.62$, characterized by a kink in the mean energy per site. Note that on the cubic lattice, our corresponding result (with \(J=0\)) is $T\sub{c} \simeq 0.71$. The low-temperature phase is always perfectly synchronized, since there is an energy $K$ for every dimer in a given replica. The high-temperature phase, which we identify with the Coulomb phase, is unsynchronized. In particular, when $k=0$, the mean energy per bond is $K/q^{2}$. This is sensible, because in this limit the replicas are independent, and from \refeq{eq:singlerootcf3} the probability of double bond occupation is $1/q^{2}$.

% Monomer on root
To confirm our identification of the high-temperature solution with the (unsynchronized) Coulomb phase, we return to the monomer expectation value $\Psi\sub{m}$ defined in \refeq{eq:gminf}. In this case, we consider the partition function with a monomer in a single replica (again on the left side of the root \(d_0\)), which is
\begin{equation}
Z\sub{m} = Z_{n}(1,0)Z_{n}(0,0) + Z_{n}(1,1)Z_{n}(0,1)\punc,
\end{equation}
while the partition function without monomers $Z$ is given by \refeq{eq:fullpf2}. From the definitions of \refeq{eq:unvnwn}, and their solution in \refeqand{eq:uvw1}{eq:uvw2}, one obtains
\beq[eq:monomerexpectationvalue]
\Psi\sub{m}=
\begin{cases}
\dfrac{\sqrt{(q-1)(q-2)(q-1-e^{-k})}}{(q-1)^{2} - e^{-k}} & \text{for \(k\ge k\sub{c}\)} \\
0 & \text{\phantom{for} \(k \le k\sub{c}\).}
\end{cases}
\eeq

The result is shown in \reffig{fig:bethelattice} (bottom panel), using the same parameters as for the mean energy per site. In the low-temperature synchronized phase (\(k < k\sub{c} < 0\)), \(\Psi\sub{m} = 0\) and the free-energy cost for an isolated monomer, \(\Delta F\sub{m} = -T\ln \Psi\sub{m}\), is infinite, while in the high-temperature unsynchronized phase, \(\Psi\sub{m} > 0\) and \(\Delta F\sub{m}\) is finite. This qualitative distinction, equivalent to the criterion based on \(G\sub{m}\) introduced in \refsec{phasediagram}, implies that the synchronization transition on the Bethe lattice is a \emph{bona fide} confinement transition.

While the model on the Bethe lattice with \(q=6\) gives a reasonable approximation to the critical temperature on the cubic lattice, it does not reproduce the correct critical behavior. This is directly evident for the heat capacity, \(\frac{\partial}{\partial T} \langle E \rangle\), which, according to \refeq{eq:energypersite2}, has a discontinuity at \(T = T\sub{c}\), as expected for a mean-field theory. For the monomer expectation value, the duality mapping to the XY model \cite{Powell2013} gives \(\Psi\sub{m} \sim t^\beta\) for \(t > 0\), where \(\beta\) is the magnetization exponent. From \refeq{eq:monomerexpectationvalue}, we find the mean-field value \(\beta = \frac{1}{2}\).

\section{Conclusions}
\label{sec:conclusions}

We have studied the phases of the double dimer model on the cubic lattice using a combination of theoretical arguments and MC simulations. Our results demonstrate the presence of a synchronization transition at a critical coupling between the replicas, which has no local order parameter but can be characterized through the confinement of monomers.

In a subsequent publication, we will address the same model on the square lattice; extensions to dimer models on other bipartite lattices are straightforward. By adapting the synchronization criterion introduced here, analogous transitions can also be expected in other systems consisting of two coupled replicas of a fractionalized phase. These include the Coulomb phase in ice models \cite{Henley2010,Castelnovo2012}, where a pair of monopoles in one replica would similarly become confined upon synchronization.

Experimental realizations of such transitions could be possible in various frustrated systems. In 2D, these include magnetic materials with a bilayer structure as well as nanomagnet arrays \cite{Nisoli2013}, which have been used to simulate ice models with a variety of geometries, constructed in a double-layer configuration. A 3D synchronization transition could be possible between magnetic moments of two types, for example, in pyrochlore oxides with magnetic ions on both the A and B sites of the crystal structure \cite{Gardner2010}. In these cases, one expects a thermodynamic phase transition (see, e.g., \reffig{fig:heatcap}), but with no magnetic ordering. We leave the detailed study of possible experimental signatures to future work.

A natural extension of the double dimer model that we have treated here is to consider the case of multiple replicas \(\alpha \in \{1,2,\dotsc,n\}\). With strong coupling between `adjacent' replicas \(\alpha\) and \(\alpha+1\), this can be interpreted as the Suzuki--Trotter decomposition of the partition function for a quantum dimer model \cite{Moessner2011}. The double-loop algorithm introduced in \refsec{doubleloops} could be extended to the case of multiple replicas, giving a method that is similar (at least in spirit) to the membrane algorithm \cite{Henry2014} previously applied to quantum ice.

%\vspace{2em}

\acknowledgments{The simulations used resources provided by the University of Nottingham High-Performance Computing Service. We are grateful to J. P. Garrahan for helpful discussions.}

\bibliography{dimersbibliography}

\end{document}